\documentclass[sigconf, nonacm]{acmart}

\usepackage{amsmath,amssymb,amsfonts}
\usepackage{algorithmic}
\usepackage{graphicx}
\usepackage{textcomp}
\usepackage{xcolor}
\def\BibTeX{{\rm B\kern-.05em{\sc i\kern-.025em b}\kern-.08em
    T\kern-.1667em\lower.7ex\hbox{E}\kern-.125emX}}
\usepackage{adjustbox}
\usepackage{pifont}

\usepackage{multirow}
\usepackage[linesnumbered,ruled,vlined]{algorithm2e}
\usepackage{subcaption}
\usepackage{colortbl}
\usepackage{array}
\usepackage{url}
\newcolumntype{K}[1]{>{\centering\arraybackslash}p{#1}}
\usepackage{todonotes}

\AtBeginDocument{%
  \providecommand\BibTeX{{%
    Bib\TeX}}}

\setcopyright{cc}
\setcctype{by}

\begin{document}

\title{Energy Metrics for Edge Microservice \\Request Placement Strategies }

\author{Klervie Toczé}
\authornote{Part of the work was performed while the author was at Linköping University.}
\email{k.m.tocze@vu.nl}
\orcid{}
\affiliation{%
  \institution{Vrije Universiteit Amsterdam}
  \country{The Netherlands}
}

\author{Simin Nadjm-Tehrani}
\email{simin.nadjm-tehrani@liu.se}
\affiliation{%
  \institution{Linköping University} 
  \country{Sweden}}

\renewcommand{\shortauthors}{Toczé and Nadjm-Tehrani}

\begin{abstract}
  Microservices are a way of splitting the logic of an application into small blocks that can be run on different computing units and used by other applications. It has been successful for cloud applications and is now increasingly used for edge applications.  
  This new architecture brings many benefits 
  but it makes deciding where a given service request should be executed (i.e. its placement) more complex as every small block needed for the request has to be placed. 
  
  In this paper, we 
  investigate energy-centric request placement for services that use the microservice architecture, and specifically whether using different energy metrics for optimization leads to different placement strategies. We consider the problem as an instance of a traveling purchaser problem and propose an integer linear programming formulation. This formulation aims at minimizing energy consumption while respecting latency requirements. 
  We consider two different energy consumption metrics, namely overall or marginal energy, when applied as a measure to determine a placement.  
  
Our simulations show that using different energy metrics indeed  results in different request placements. The paper presents several parameters influencing the extent of this difference.  
\end{abstract}

\ccsdesc[500]{Computer systems organization~Cloud computing}
\ccsdesc[500]{Networks~Cloud computing}
\ccsdesc[500]{Networks~Network resources allocation}

\keywords{Energy metrics, optimization, edge/fog computing.}

\maketitle

\section{Introduction}

The edge computing paradigm, consisting of moving computational and storage resources to the edge of the network, is envisioned to obtain  lower latencies, increase privacy and alleviate the amount of data sent to a distant cloud. 
As edge computing infrastructures are being deployed, the microservice architecture is subject to intensive study, both in the cloud and at the edge. With this architecture, services are decomposed into a chain of different functions, allowing for higher flexibility and sharing of the function logic between different services.

We consider a distributed edge infrastructure 
where each edge device receives service requests coming from end users through end devices. An end device can e.g. be an IoT device (a video surveillance camera, a sensor), a mobile phone, a connected vehicle, etc..
The edge computing infrastructure is composed of heterogeneous resource-limited devices. This means that functions cannot deploy function instances on all available edge devices as this will be too resource-hungry and also because some specific function instances can only run on specific edge hardware. For example, some machine learning algorithms require GPU resources to run.

In this work, we study microservice \textbf{request} placement, i.e. for a given set of function instances already deployed in the edge infrastructure, which one(s) to select for executing a particular request, incoming at a given physical location and at a given point in time. 
How to best deploy the function instances for improved performance and how many of them should be deployed are related placement problems which are tackled by other works~\cite{Tocze_Violinn, Rubak_PredictiveResourceScaling}. 
The request placement problem is critical to be addressed since it is where the demand side (the application requirements) and the supply side (the infrastructure provisioning) meet. 
The application provider is interested in the requests completing in time to guarantee a high quality of experience for its users. The edge infrastructure provider is interested because the request placement can be used to improve load balancing 
or  load consolidation.

Providing (micro)services using edge computing consumes energy. As part of an effort to minimize the impact of edge computing on energy consumption, we want to study request placements with an energy-centric perspective. Indeed, a common optimization objective, as many edge applications are latency-critical, is to minimize the completion time of a service request. Instead, we argue that there is no benefit in minimizing the service completion time as long as it is below the deadline requirement, i.e. that its responsiveness is good enough. Therefore, we use the deadline as a constraint and choose to optimize with regards to energy consumption only. We formulate two different energy metrics with different underlying ideas: one considers the energy consumption as traditionally done and the other considers the energy consumption increment created by the current request being placed. 
Our goal is to investigate whether the choice of energy metric results in different placement decisions and if so, what are the characteristics of the difference obtained and which factors  influence it. 
As an aid in this investigation, we formulate the microservice request placement problem as an instance of the Traveling Purchaser Problem (TPP). The corresponding optimization problem is solved using Integer Linear Programming (ILP). 

The contributions of this paper are the following:
\begin{itemize}
    \item We formulate two energy metrics  leading to two energy-centric placement strategies. 
    \item We express the microservice request placement problem as a Traveling Purchaser Problem and propose an ILP formulation for it, using the two formulated energy metrics for the two different optimization objectives. 
    \item We analyze the obtained placements and study the impact of the two metrics on the placement decision under varying load and other system parameters. 
\end{itemize}

The rest of the paper is organized as follows: in Section \ref{sec:Notations} we introduce the system model considered. Section \ref{sec:EnergyConsumption} focuses on energy and describes the models used and the proposed metrics. Section \ref{sec:Problem} presents the problem and its two formulations. 
The evaluation performed is described in Section \ref{sec:Evaluation}.  
The outcomes are presented in Section \ref{sec:EvaluationOutcomes}, including a study of the influence of varying different system characteristics on the placement decision. We discuss related works in Section \ref{sec:RelatedWorks}, and conclude in Section \ref{sec:Conclusion}. 

\section{System model}
\label{sec:Notations}

\subsection{Edge infrastructure}

The edge devices included in the considered part of the edge infrastructure are represented as a directed graph\footnote{We choose not to use the classic E notation for edges to avoid confusion with the edge of edge computing. Instead we use an L for "Links". } $G=(V,L)$.
Each vertex of the graph corresponds to an edge device and each link of the graph corresponds to a communication link between two edge 
devices. 
Thus, we assume we have information about which edge devices are in the system and how they are connected. 
Aditionally, such information may include the device current utilization level, its performance profile,  and its energy profile. Information about the links may include their propagation delay and transmission power. 
How this information is retrieved and where/how often this retrieval takes place is out of the scope of this paper. 

\subsection{Service}

The end users want to use various \textit{services}. For example, an augmented reality application is used to visualize an architectural project, or the video footage of a crossing needs to be analyzed.

In this paper, the considered services are following the microservice architecture, i.e. "an approach for developing a single application as a suite of small services, each running in its own process and communicating with lightweight mechanisms"~\cite{Lewis_MicroserviceDef}. 
Therefore, a service $S$ can be defined as a directed acyclic graph $S=(F,D)$ where a vertex $f \in F$ is  one of the functions (micro services) the service is composed of and an edge $d \in D$ is a dataflow between two different functions~\cite{Li_ServerlessComputing_Survey}. 
A service $S$ is therefore a chain of functions connected with a sequence of edges in the same direction.
A function $f\in F$ is associated with a computing size $f^s$ and a dataflow $d \in D$ is associated with a data size $d^s$.  

In order to provide a service $S$, function instances for every function $f \in F$ are deployed on  different edge devices $v$ present in the infrastructure. At most one function instance for a given function $f$ is deployed on a given node $v$. We therefore introduce $\phi=(v,f)$ as a shorthand for describing the function instance for function $f$ deployed on node $v$. 
The function instances can then handle incoming requests. 

To execute a function which is a part of a service, an edge device needs to have the necessary hardware and software available. This is why all functions may not execute on all edge devices. The necessary software is for example provided using a dedicated container~\cite{Oleghe_ContainerPlacementAndMigration}. 
We denote $V_f \subseteq V$ the subset of edge devices able to execute a given function $f$, i.e. the subset of edge devices having a function instance for function $f$ deployed.  
We consider that each function is available on a subset of edge devices only ($\forall f, V_f \subset V$) and that every function is available on at least one edge device ($\forall f, V_f\neq \emptyset$).

\subsection{Service request}

The load coming to the edge infrastructure consists of service requests. A service request $r=(S,v_b,v_e,\theta, \delta) \in R$ is defined by:
\begin{itemize}
    \item the requested service $S$
    \item the edge device receiving the request (beginning device) $v_b \in V$
    \item the edge device receiving the request answer (end device) $v_e \in V$ 
       \item the request arrival time $\theta$ 
    \item the request deadline $\delta$
\end{itemize}

The deadline $\delta$ is relative to the service request arrival time $\theta$ and corresponds to the maximum time allowed for the request to go through all the functions comprising the service and reach the destination device.  
In this work, we consider services for which it is necessary to complete requests before the deadline, otherwise their quality of service (QoS) is severely degraded.

\subsection{Request placement}

In this paper, we focus on the request placement problem. That is, for each request $r$ with service $S$ incoming to the edge system, the edge system has to decide on  a placement $p$ so that the request $r$ will be completed before its deadline $\delta$. We assume that 1) the edge devices have been provisioned and 2) a set of function instances has been placed on various edge devices in order to be able to provide for the services asked by the end users. 

When handling a given service request, deciding on which function instance (and by extension on which edge device) each of the functions of a service $S$ will be executed is called placing the request. Such a service request placement is denoted $p=(\phi_1,...,\phi_{|F|})$ where $\phi_i$ corresponds to the function instance chosen for executing the $i^{th}$ function of service $S$. 

For a given placement $p$,  the set of edge devices included in the placement $V_p\subseteq V$ (i.e. the ones where the selected function instances are deployed) is denoted by $V_p = \{v | \phi=(v,f) \in p\}$. $L_p\subseteq L$ is the set of links connecting these edge devices.

In this work, we focus on request placement taking place at one decision point at a time, with a specific current resource utilization.  
 The aim is to be able to study how different optimization objectives will influence the placement decision.

\subsection{Request completion time}
\label{sec:requestCompletionTime}
Once the placement $p$ of a service request $r$ is known, it is possible to calculate the completion time of the request in order to see whether it is below the deadline $\delta$. 

The completion time of a request is composed of two parts:
\begin{itemize}
    \item The transmission time
    \item The execution time.
\end{itemize}

The transmission time depends on the link propagation delay (in ms), the available link bandwidth (in byte/ms), and the size of the data (in byte) that needs to be transferred on the link. 
We assume that the request utilizes all the available link bandwidth to transmit. 
Therefore, the transmission time of dataflow $d$ with data size $d^s$ on a given link $l$ is:
\begin{equation}
\label{eq:linkCompletionTime}
    \lambda_{ld}=l^l+\frac{d^s}{l^c}
\end{equation}
where $l^l$ is the propagation delay for link $l$, and $l^c$ is the currently available bandwidth for the link $l$. 
We assume that the link(s) chosen correspond to the shortest path between the two edge devices.

The execution time depends on the available computing capacity of the edge device (in million of instruction (MI) per millisecond) and the size of the function needing to be computed (in MI). 
Each function instance running on the edge device gets a share of the full capacity. This share (i.e. the available computing capacity for the function instance) can account e.g. for the need to always have some free capacity and thus may vary over time.
We assume that the request utilizes all the available computing capacity in the corresponding function instance for the function execution. 
The execution time of the function instance $\phi=(v,f)$ corresponding to function $f$ with computing size $f^s$ deployed on edge device $v$ is therefore calculated as follows:
\begin{equation}
\label{eq:sys_executionTime}
     \lambda_{\phi}=\frac{f^s}{\phi^c}
\end{equation}
 where $\phi^c$ is the available computing capacity allocated to the function instance $\phi$. 

The total completion time of a service request $r$ to service $S=(F,D)$ using a given placement $p$ is the sum  of the transmission times and execution times for all link transfers and function executions required to complete the service request.
This can be expressed as:
\begin{equation}
\begin{split}
    \label{eq:latencyTotal}
    \Lambda_{rp}=\sum_{l \in L_p}\sum_{d \in D}\mathcal{I}^{p}_{ld}*\lambda_{ld}+\sum_{v \in V_p}\sum_{f \in F}\mathcal{I}^{p}_{vf}*\lambda_{\phi} \\
    \text{where} \qquad \mathcal{I}^{p}_{ld}\begin{cases}
         1, & \text{if $d$ is sent over $l$ according to $p$}\\
            0, & \text{otherwise.}
            \end{cases}
    \\
    \text{and} \qquad \mathcal{I}^{p}_{vf}\begin{cases}
         1, & \text{if $f$ executes on $v$ according to $p$}\\
           0, & \text{otherwise.}
            \end{cases}
\end{split}
\end{equation}

\section{Energy-centric placement}
\label{sec:EnergyConsumption}

This work places the focus on the energy footprint of placement decisions and considers that lowering the energy consumption should be the sole optimization objective, with performance being a constraint. We want to study whether considering different energy metrics to optimize for can result in different placement decisions. In this section, we present the models used as well as the studied energy metrics and associated envisioned strategies. 

\subsection{Energy models}
\label{sec:EnergyModels}

Following Baccarelli et al.~\cite{Baccarelli_EcoMobiFog} and Ahvar et al.~\cite{Ahvar_EstimatingEnergy}, we model the energy consumed by an edge device or an edge link as having both a static part (corresponding to the device/link being in the idle state) and a dynamic part (corresponding to the energy needed for processing/transmitting).

\subsubsection{Edge links}

The energy used by an edge link $l$  to service a request through the transmission of the dataflow $d$ is calculated as follows:
\begin{equation}
\label{eq:energyLink}
    E_{ld}=\underbrace{\mathcal{P}^{IDLE}_l*\lambda_{ld}}_{\text{static part}}+\underbrace{\mathcal{P}^{DYN}_l*\lambda_{ld}}_{\text{dynamic part}}
\end{equation}
where $\mathcal{P}^{IDLE}_l$  is the idle power needed for maintaining the link $l$ (e.g. the power consumed by the NIC cards at both ends when in the idle state) and $\mathcal{P}^{DYN}_l$ is the sum of the power needed by the link $l$ for transmitting from the transmitting node and receiving at the receiving node. These are related to  a wide range of characteristics of the communication link (e.g. number of antennas) and the current link throughput~\cite{Baccarelli_EcoMobiFog}. $\lambda_{ld}$ is the duration of using $l$ for transmitting $d$, according to Section \ref{sec:requestCompletionTime}. 

\subsubsection{Edge devices}

The energy consumption for an edge device $v$ used to service a request  through the execution of the function $f$ is:
\begin{equation}
\label{eq:energyDeviceModel}
    E_{\phi}=\underbrace{\mathcal{P}^{IDLE}_v*\lambda_{\phi}}_{\text{static part}}+\underbrace{\mathcal{P}^{DYN}_v(u_{\phi})*\lambda_{\phi}}_{\text{dynamic part}}
\end{equation}
where  $\mathcal{P}^{IDLE}_v$ is the power needed for device $v$ to be on and $\mathcal{P}^{DYN}_v(u_{\phi})$ is the extra power needed for device $v$ for executing at a utilization level of $u_{\phi}$ (that includes the execution of the function instance $\phi$). $\lambda_{\phi}$ is the duration of executing $f$  on $v$, according to Section \ref{sec:requestCompletionTime}.

To model the dynamic power, we use the piecewise-linear model proposed by Ahvar et al.~\cite{Ahvar_EstimatingEnergy} with their measurements for the Parasilo server (illustrated in Figure \ref{fig:powerProfileParasilo}). Hence, the dynamic power is written as $\mathcal{P}^{DYN}_v(u_{\phi})=(P^{cores}_{j+1}-P^{cores}_{j})* k* u_{\phi}+[(j+1)* P^{cores}_{j}-j* P^{cores}_{j+1}]$ where $j/k \leq u_{\phi}\leq (j+1)/k$ with $j\in \{0,...,k-1\}$ where $k$ is the total number of cores that the devices has, and $P^{cores}_{j}$ is the dynamic power consumption value when j cores are utilized.

Note that $\mathcal{P}^{IDLE}_v$ is the idle power of the \textit{full} edge device. 
This corresponds to the worst case where only one function instance 
is deployed. 
If several function instances are deployed, one can define variants of Equation \ref{eq:energyDeviceModel}  where only an apportioned chunk of the idle power is taken into account as the energy consumption of a given instance. Determining this chunk can be done using energy apportionment~\cite{Vergara_Apportionment}. Such variants are out of the scope of this work. 

\subsection{Energy metrics}
\label{sec:EnergyMetrics}

We introduce two different energy metrics for calculating the energy consumption associated with executing a function instance $\phi$:
\begin{enumerate}
    \item The overall energy consumption, i.e. how much the device will consume after placing the execution of $f$ on top of what it is already executing.
    \item The marginal energy consumption, i.e. how much additional energy does the execution of function $f$ consumes on the device.
\end{enumerate}

\subsubsection{Associated envisioned placement strategies}

Each of the above two metrics is associated with an envisioned  placement strategy. 

For the overall energy metric, the idea is to place the requests on the devices that will lead to the lowest energy consumption, when looking at the total energy consumption of all devices involved in the placement. This is inline with traditionally used energy consumption metrics. 

For the marginal energy metric, the idea is to favor load consolidation, i.e. to avoid putting load on a device which is currently idle. Load consolidation makes it possible to gather the load on a few devices and switch off the other ones instead of having more devices on but at a low utilization. 

For both metrics, the idea behind considering the response time as a constraint only and not an optimization objective, is to allow for placement on devices that require some additional communication but offer better characteristics with regards to the target energy metrics than closer devices. 

\subsubsection{Overall energy}

The overall energy consumption for an edge device $v$ used to service a request  through the execution of the function $f$ is therefore:
\begin{equation}
    E^O_{\phi}=E_{\phi} \qquad \text{as defined in Equation \ref{eq:energyDeviceModel}}
\end{equation}

\subsubsection{Marginal energy}

The marginal energy consumption for an edge device $v$ used to service a request  through the execution of the function $f$ depends on what was the utilization of device $v$ before the execution of $f$ ($u_v$). It is calculated as:
\begin{equation}
\label{eq:energyDevice}
    E^M_{\phi}=\begin{cases}
         E^O_{\phi}, & \text{if $u_v =0$ }\\
            \mathcal{P}^{DYN}_v(u_{\phi}-u_{v})*\lambda_{\phi}, & \text{otherwise.}
    \end{cases}
\end{equation}

The difference between the overall and marginal energy consumption for a given execution of a function instance represents whether we need to use a previously unused device for the execution (first line in Equation \ref{eq:energyDevice}) or whether the function is added to an already used device. This is useful to know because if the static energy consumption represents an important part of the energy consumption, then an energy-efficient placement should favor the devices already in use instead of turning on new ones. 

\subsection{Request energy consumption}

The request energy consumption, in a similar way to the request completion time (see Section \ref{sec:requestCompletionTime}), has two parts: 
\begin{itemize}
    \item The transmission energy consumption
    \item The execution energy consumption.
\end{itemize}

Given the models from Section \ref{sec:EnergyModels} and the two metrics from Section \ref{sec:EnergyMetrics}, the total overall/marginal energy consumption of a service request $r$ to service $S=(F,D)$ using a given placement $p$ is obtained by summing the corresponding energy consumption of all dataflow transmissions and the corresponding energy consumption for computing each function on the edge devices included in the placement solution. 
This is expressed as:
\begin{equation}
\begin{split}
    \label{eq:energyTotal}
    E^O_{rp}=\sum_{l \in L_p}\sum_{d \in D}\mathcal{I}^{p}_{ld}E_{ld}+\sum_{v \in V_p}\sum_{f \in F}\mathcal{I}^{p}_{vf}E^O_{\phi}
    \\
   \text{and} \quad E^M_{rp}=\sum_{l \in L_p}\sum_{d \in D}\mathcal{I}^{p}_{ld}E_{ld}+\sum_{v \in V_p}\sum_{f \in F}\mathcal{I}^{p}_{vf}E^M_{\phi}
    \\
    \text{where} \qquad \mathcal{I}^{p}_{ld}\begin{cases}
         1, & \text{if $d$ is sent over $l$ according to $p$}\\
            0, & \text{otherwise.}
            \end{cases}
    \\
    \text{and} \qquad \mathcal{I}^{p}_{vf}\begin{cases}
         1, & \text{if $f$ executes on $v$ according to $p$}\\
           0, & \text{otherwise.}
            \end{cases}
    \end{split}
\end{equation}

\section{Problem formulations}
\label{sec:Problem}

In this work, the request placement problem is solved for each individual request coming to the edge. 

\subsection{TPP formulation}
The service request placement problem can be expressed as an instance of the Traveling Purchaser Problem (TPP)~\cite{Manerba_TPPandVariants}. 

The TPP is a generalization of the well-known Traveling Salesman Problem (TSP), where there are different marketplaces that sell a given set of items at a given price. The problem is defined for a purchaser that has a given list of items to buy, to find the route between the marketplaces that minimizes both the cost of travel and purchase. 

In our case, the purchaser corresponds to the service request that has to travel to different edge devices (marketplaces) offering the functions (items) composing the service requested (the list of items). 
The approach is energy-centric and considers the costs of travel and purchase to be the energy consumption of using a link or an edge device, according to their descriptions in Section \ref{sec:EnergyConsumption}. 

In order to adopt this approach, we need to add two constraints to the generic formulation of the TPP. The first one is that there is a given order in which the functions should be executed, so that the service is executed properly. Secondly, it is not enough to minimize the cost of the request placement, the placement should also meet the request deadline to be an acceptable solution. 

\subsection{ILP formulation}
An optimal solution to the service request placement problem can be found using integer linear programming (ILP). 
This section details the integer linear programming (ILP) formulation used to obtained placement decisions in this work. 
Table \ref{tab:notationILP} summarizes the notations used.

\begin{table}[]
    \centering
    \begin{tabular}{|K{2cm}|p{0.7\columnwidth}|}
    \hline
        \textbf{Symbol} & \textbf{Meaning} \\
        \hline
       
        $x_{\phi\psi}$& Decision variable indicating whether the link between function instances $\phi$ and $\psi$ is included in the solution\\\hline
        $y_{\phi}$& Decision variable indicating whether  the function instance $\phi$ is included in the solution \\\hline
        $o_{\phi\psi}$& Decision variable indicating the order in which the link between function instances $\phi$ and $\psi$ is visited in the solution \\\hline
        $\Phi 
        $& Set of function instances\\\hline
                  $F= (f_1,f_2,....,f_{|F|})$& Ordered list of service functions to be placed\\\hline 
         $\phi_b$& Beginning function instance located on device $v_b$\\\hline
         $\phi_e$& End function instance located on device $v_e$\\\hline
         $E_{(\phi,\psi)}$& Energy consumption of the link between any two function instances $\phi$ and $\psi$ (including zero for modelling both being on the same device)\\\hline
         $E_{\phi}$& Energy consumption of the function instance $\phi$ \\\hline
         $a^{\phi}_f$& Boolean indicating whether function instance $\phi$ is an instance of function $f$ (=1) or not (=0). \\\hline
         $O_{f}=i$& Positive integer representing the position of the function $f$ in the list $F$ \\\hline
         $\lambda_{\phi\psi}$& Transmission time for the link between the function instances $\phi$ and $\psi$\\\hline
         $\lambda_{\phi}$& Execution time for the function instance $\phi$\\\hline
         $\beta$& Integer upper bound used when ordering the visited links. It should be greater that the number of links contained in the service chain. 
         \\\hline
    \end{tabular}
    \caption{ILP notations.}
    \label{tab:notationILP}
\end{table}

\subsubsection{The infrastructure graph}

In order to enforce the ordering constraint between the different functions composing the service, we use a variable assigning an integer to the links that are part of the placement (in a similar way to the work by Shameli-Sendi et al.~\cite{Shameli-Sendi_EfficientProvisioning}). In order to both enable a placement to have several service instances on the same physical node (good for load consolidation) and to allow the same edge device to be selected for different (non-consecutive) functions, 
we consider a device in the ILP infrastructure graph to be  a function instance and not a hardware device. Consequently, a link is a virtual link between function instances, that can be mapped to a physical one.
We also represent the beginning and end devices ($v_b$ and $v_e$) as two specific "virtual" function instances denoted $\phi_b$ and $\phi_e$ respectively.  

Note that since the nodes in the ILP graph are function instances, it is not necessary to specify both the edge device and the function (as a function instance is the combination of both) where referring to a node. This means that for the completion time and energy consumption equations (Equations \ref{eq:linkCompletionTime} to \ref{eq:energyTotal}), the subscript $ld$ can be replaced by the corresponding start and end function instances. 

\subsubsection{Decision variables}

This formulation uses three different decision variables. 
$x_{\phi\psi}$ indicates whether the link between function instances $\phi$ and $\psi$ is selected to be part of the solution.
$y_{\phi}$ indicates whether the function instance $\phi$ is included in the placement. 
$o_{\phi\psi}$ captures the order of visiting the links in the solution. If the link $(\phi,\psi)$ is not visited, then $o_{\phi\psi}=0$. If the link $(\phi,\psi)$ is visited before the link $(\chi,\zeta)$ then we have  $o_{\phi\psi} > o_{\chi\zeta}$.
In case the link $(\phi,\psi)$ is visited just before the link $(\chi,\zeta)$ then we have  $o_{\phi\psi} = o_{\chi\zeta}+1$.

\subsubsection{Objective function}
\label{sec:objectiveFunction}

The objective function in this work is focusing on energy consumption only. The performance requirements (here that a request should complete before a deadline), are considered as a constraint instead of an optimization objective. 
The energy consumption metric considered here is either the overall or the marginal one described in Section \ref{sec:EnergyConsumption}. 
This work is focusing on minimizing resource use, therefore the objective function is: 
$$\text{Minimize } \sum_{\phi \in \Phi}\sum_{\psi \in \Phi} E_{\phi\psi}* x_{\phi\psi} + \sum_{\phi \in \Phi\setminus\{\phi_b,\phi_e\}} E_{\phi}*y_{\phi}$$

The device energy consumption metric $E_\phi$ in the equation above is replaced by the overall ($E^O_\phi$) or the marginal ($E^M_\phi$) energy consumption metric depending on the one chosen for optimizing. 

\subsubsection{Constraints}
The ILP formulation contains the following constraints:
\begin{equation}
    \label{eq:S_edge}
    \sum_{\psi \in \Phi} x_{\psi\phi} = \sum_{\zeta \in \Phi} x_{\phi\zeta} \qquad\forall \phi \in \Phi\setminus\{\phi_b,\phi_e\}
\end{equation}
\begin{equation}
    \label{eq:S_edge_start}
    \sum_{\psi \in \Phi} x_{\psi\phi} + 1 = \sum_{\zeta \in \Phi} x_{\phi\zeta} \qquad  \phi = \phi_b
\end{equation}
\begin{equation}
    \label{eq:S_edge_final}
    \sum_{\psi \in \Phi} x_{\psi\phi}  = \sum_{\zeta \in \Phi} x_{\phi\zeta} + 1 \qquad \phi = \phi_e
\end{equation}

\begin{equation}
    \label{eq:S_quantity}
     \sum_{\phi \in \Phi\setminus\{\phi_b,\phi_e\}} (y_{\phi}*a_f^{\phi}) = 1 \qquad\forall f \in F
\end{equation}
\begin{equation}
 \label{eq:S_include}
     \sum_{\psi \in \Phi} x_{\phi\psi} -y_{\phi} = 0  \qquad\forall \phi \in \Phi\setminus\{\phi_b,\phi_e\} 
\end{equation}

\begin{equation}
 \label{eq:S_avoid1}
    \sum_{\phi \in \Phi} x_{\phi\psi} \leq 1  \qquad \forall \psi \in \Phi
\end{equation}
\begin{equation}
 \label{eq:S_avoid2}
    \sum_{\psi \in \Phi} x_{\phi\psi} \leq 1  \qquad \forall \phi \in \Phi
\end{equation}

\begin{equation}
 \label{eq:S_visit}
    o_{\phi\psi} \leq \beta * x_{\phi\psi}  \qquad \forall \phi \in \Phi,\forall \psi \in \Phi
\end{equation}
\begin{equation}
 \label{eq:S_calcul}
\sum_{\psi \in \Phi}o_{\psi\phi} = \sum_{\psi \in \Phi}(o_{\phi\psi} +x_{\phi\psi}) \qquad \forall \phi \in \Phi \setminus \{\phi_b,\phi_e\}
\end{equation}
\begin{equation}
 \label{eq:S_calcul_source}
\sum_{\psi \in \Phi}o_{\psi\phi} + \beta = \sum_{\psi \in \Phi}(o_{\phi\psi} +x_{\phi\psi}) \qquad \phi =\phi_b 
\end{equation}

\begin{equation}
 \label{eq:S_order}
 \begin{aligned}
     (\sum_{\alpha\in \Phi\setminus \{\phi_e\}} o_{\alpha\phi}) -x_{\phi\psi} - \beta*y_{\phi} +\beta \geq \\ (\sum_{\alpha\in \Phi\setminus \{\phi_e\}} o_{\alpha\psi}) - \beta * y_{\psi} +\beta * \sum_{\alpha\in \Phi\setminus \{\phi_e\}} x_{\alpha\psi}  \\ 
     \forall \phi,\psi \in \Phi \setminus \{\phi_b,\phi_e\}, \phi \ne \psi, \\ \forall f,k \in F, f \ne k, O_{f} = O_{k} -1, a^{\phi}_f=a^{\psi}_k=1
 \end{aligned}
\end{equation}

\begin{equation}
 \label{eq:S_deadline}
    \sum_{\phi \in \Phi}\sum_{\psi \in \Phi}\lambda_{\phi\psi}*x_{\phi\psi}+\sum_{\phi \in \Phi \setminus \{\phi_b,\phi_e\}}\lambda_{\phi}*y_{\phi} \leq \delta \qquad  
\end{equation}

\begin{equation}
 \label{eq:S_noSelfLoop}
    \sum_{\phi \in \Phi}x_{\phi\phi}=0  
\end{equation}

\begin{equation}
 \label{eq:S_binX}
     x_{\phi\psi} \in \{0,1\}  \qquad\forall \phi,\psi \in \Phi
\end{equation}
\begin{equation}
 \label{eq:S_binY}
     y_{\phi} \in \{0,1\}  \qquad\forall \phi \in \Phi \setminus \{\phi_b,\phi_e\}
\end{equation}
\begin{equation}
 \label{eq:S_binV}
    o_{\phi\psi}\in \mathbb{N}, \qquad 0 \leq o_{\phi\psi} \leq \beta,    \qquad\forall \phi,\psi \in \Phi 
\end{equation}

Constraint \ref{eq:S_edge} ensures that for any function instance $\phi$, there is one edge going in and out the function instance. For the special cases of the source (Constraint \ref{eq:S_edge_start}) and destination (Constraint \ref{eq:S_edge_final}), a virtual incoming (respectively outgoing) edge is added to them.
Constraint \ref{eq:S_quantity} ensures that only one  function instance is selected per function to be placed and that the function instance selected can actually run this function. 
Constraint  \ref{eq:S_include} ensures that if a function instance is chosen to be included in the solution, then it has to be on the solution path. 
Constraints  \ref{eq:S_avoid1} and \ref{eq:S_avoid2} ensure the solution does not include any cycle. This is needed since the ordering constraint can only assign one order number to each link, hence cycles are impossible. 
Constraint  \ref{eq:S_visit} ensures that if a link is not selected, the value of the corresponding $o_{\phi\psi}$ is set to 0. If a link is selected, the corresponding ordering variable $o_{\phi\psi}$ is less than $\beta$. 
Constraint  \ref{eq:S_calcul} is used to calculate the values of $o_{\phi\psi}$ for each selected link.
The special case of the start node is covered in Constraint \ref{eq:S_calcul_source}.
Constraint  \ref{eq:S_order} ensures that the functions are placed in the indicated order. The order is specified by $O$, where if $O_l = O_{k}-1$, it means that the function $l$ is placed before function $k$ and that the function instance $\phi$ where $y_{\phi}*a^{\phi}_l=1$ is visited before the function instance $\psi$ where $y_{\psi}*a^{\psi}_k=1$. 
Constraint \ref{eq:S_deadline} ensures that the latency corresponding to the selected path (i.e. a set of function instances connected through ordered links) is below the latency requirement, i.e. the deadline of the service request.
Constraints \ref{eq:S_binX} and \ref{eq:S_binY} indicate that the decision variables are binary. Constraint \ref{eq:S_binV} indicates that the decision variable is a positive integer. 

\section{Evaluation setup}
\label{sec:Evaluation}

The proposed ILP formulation is implemented and this section describes how its evaluation is conducted. 

\subsection{Edge system}
\label{sec:eval_edgeSystem}
The edge system considered in this evaluation has the same topology as the Abilene network~\cite{Abilene}. The distances between the edge devices are scaled to correspond to a scenario where all the edge devices are spread within a neighborhood of a city. The propagation delay between two edge devices is taken as proportional to the distance between these two devices. This model can easily be adapted to consider different delay characteristics (e.g. link propagation depending on connection technology). 

A service request arrives at the closest edge device. For placing it, some knowledge about the current state of the system is required, such as which function instances are available and current utilization levels. How this information is obtained and updated (and how accurate it is) is out of the scope of this paper. 

All the edge devices and links in the system have the same characteristics with regards to processing/transmitting capacity, as well as the same energy profile. A summary of the system characteristics is shown in Table \ref{tab:system}.  The energy values are taken from the measurements performed by Ahvar et al.~\cite{Ahvar_EstimatingEnergy}. 

\begin{table}[]
    \centering
    \begin{tabular}{|c|c|}
    \hline
        \textbf{Characteristic} & \textbf{Value}  \\
        \hline
         Edge device computational capacity& 500 MI/ms \\
         \hline
         Edge device number of cores& 16 \\
        \hline
        Edge device idle power& 98 W \\ 
        \hline
        Edge device dynamic power at full load& 143 W \\ 
        \hline
        Link bandwidth & 500 MB/ms \\
        \hline
        Link idle power &  1 W\\ 
        \hline
        Link dynamic power &  9 W\\ 
        \hline
               
    \end{tabular}
    \caption{System characteristics}
    \label{tab:system}
\end{table}

\subsection{Service} 
The edge service considered in this evaluation is composed of four functions that are executed sequentially. 
Table \ref{tab:application} summarizes its different characteristics.

We consider that the same device that issues the request waits for the result, meaning that 
the execution result has to be transmitted back  to the beginning device after the service request execution is completed. 
An example of such a service is a mixed reality application where service requests contain images that have to be decoded (F1), analyzed to create a virtual representation of the scene (F2), modified with the addition of virtual elements (F3) and encoded (F4), before being sent back to the initial (issuing) device for rendering to the end user~\cite{Tocze_JoCC}.

\begin{table}[]
    \centering
    \begin{tabular}{|c|c|c|}
    \hline
        \textbf{Characteristic} & \multicolumn{2}{c|}{\textbf{Value}}  \\
        \hline
         Request deadline& \multicolumn{2}{c|}{100 ms} \\
        \hline
        \multirow{4}{*}{Computation} & Function 1 (F1) & 20 MI \\
        \cline{2-3}
        & Function 2 (F2) & 200 MI \\
        \cline{2-3}
        & Function 3 (F3)& 200 MI \\
        \cline{2-3}
        & Function 4 (F4)& 20 MI \\
        \hline
        \multirow{5}{*}{Communication} & Dataflow 1 & 250 MB \\
        \cline{2-3}
        & Dataflow 2 &  500 MB\\
        \cline{2-3}
        & Dataflow 3 & 750 MB \\
        \cline{2-3}
        & Dataflow 4 & 500 MB \\
         \cline{2-3}
        & Dataflow 5 & 250 MB\\
        \hline
        
    \end{tabular}
    \caption{Service characteristics}
    \label{tab:application}
\end{table}

Moreover, the number of function instances for each function and their localization is predefined. 
We recall that where to place them and how many of them should be placed 
is out of the scope of this paper. 
In this evaluation, two function instances are initially deployed for each of the functions on two different edge devices and the edge devices chosen are different for the different functions.  

\subsection{Load scenario}
\label{sec:loadScenario}

The load in the evaluation consists of different utilization levels for the edge devices and communication links. These utilization levels indicate which share of the total computation/communication capacity is available for the request being placed. Each function instance gets up to one core of computing capacity.  
For this evaluation, the utilization levels (in percentage) are decided randomly using a Normal distribution. 
Initially, we take a distribution with a standard deviation of 10.

\subsection{Evaluation approach}
\label{sec:Simulator}

The evaluation consists of groups of 40 runs each. Each run corresponds to a different random load scenario (as described in Section \ref{sec:loadScenario}). 
A group corresponds to a given load distribution and a given number of instances available for each function. A summary of the twelve different group types considered is presented in Table \ref{tab:evaluation_groups}. 

The columns present the characteristics of each group type, starting with an identifier (column \#) and a short name (column Name). The next two columns describe how the load for the edge devices, respectively the communication links is generated (columns load distribution edge devices/communication links). Then the number of function instances available for each function in the evaluation is shown (column \# instances per function). The last column indicates where the results from the given group types are presented. 

The rows present the different group types. Row 1 (Initial) provides the scenario for an initial evaluation of the obtained placement solutions. Rows 2 to 4 are studying different load distributions for the edge devices. In rows 2 and 4, the load level on each edge devices is obtained using a normal distribution, with a standard deviation of 10 in row 2 and 30 in row 4. In row 3, the load level is the same for all edge devices and is fixed to a value between 0 and 100, increasing by steps of 10. Rows 5 and 6 are studying the impact of having more function instances available (4 for row 5 and 6 for row 6). The other parameters are the same as row 2. Row 7 studies a specific case where the beginning device can execute all functions. Its parameters are the same as row 6 otherwise. Row 8-12 repeats the study from rows 2 to 6 but the beginning device is randomly varying instead of being fixed. 

When the load distribution edge devices column contains a cell with the mention "varying-load-0-100", the evaluation is repeated with the load level varying between 0 and 100 in steps of 10, the other parameters being the same. Hence, the corresponding row of the table contains 11 groups of runs.
\begin{table*}[]
    \centering
    \begin{tabular}{|c|c|K{5cm}|K{3.35cm}|K{1.9cm}|c|}
    \hline
         \textbf{\#}&\textbf{Name}&\textbf{Load distribution edge devices}&\textbf{Load distribution communication links} &\textbf{\# instances per function}& \textbf{Presented in} \\ \hline
        1&Initial& $\mathcal{N}$(varying-load-0-100, 10) &$\mathcal{N}$(varying-load-0-100, 10) &2& Section \ref{sec:results} \\ \hline
        2&Normally distributed load& $\mathcal{N}$(varying-load-0-100, 10) &No load &2& Section \ref{sec:influenceHeterogeneity}\\\hline
        3&Fixed load& Fixed load level at a value between 0-100 &No load&2& Section \ref{sec:influenceHeterogeneity} \\\hline
        4&Larger standard deviation& $\mathcal{N}$(varying-load-0-100, 30)&No load&2& Section \ref{sec:influenceHeterogeneity} \\\hline
        5&4 function instances&$\mathcal{N}$(varying-load-0-100, 10) &No load&4& Section \ref{sec:InfluenceAvailableInstances}\\\hline
        6&6 function instances&$\mathcal{N}$(varying-load-0-100, 10) &No load&6&Section \ref{sec:InfluenceAvailableInstances} \\\hline
        7&Full co-location&$\mathcal{N}$(varying-load-0-100, 10) &No load&6&Section \ref{sec:FullColocation} \\\hline
        8-10&Random beginning device&Fixed load level at a value between 0-100/$\mathcal{N}$(varying-load-0-100, 10/30) &No load&2&Section \ref{sec:randomRequesting} \\\hline
        11-12&Random beginning device&$\mathcal{N}$(varying-load-0-100, 10) &No load&4/6&Section \ref{sec:randomRequesting} \\\hline
    \end{tabular}
    \caption{Summary of the performed run group types.}
    \label{tab:evaluation_groups}
\end{table*}

In all runs of all groups, we place one request using our ILP formulation, both when minimizing overall energy consumption and also marginal energy consumption. These are named Overall and Marginal for short. The beginning device for the request is the same for the whole group, except for groups 8-12.  In total there are 10560 runs executed. 

The evaluation is performed on a Surface Laptop 5 equipped with an Intel Core i7-1265U CPU (2.7 GHz) and 16GB of RAM. Version 11.0.3 of the Gurobi Optimizer 
is used as the optimization solver. 
The code used for running the evaluation as well as the result files are provided open-source in Gitlab\footnote{\url{https://gitlab.liu.se/ida-rtslab/public-code/2024_energy-aware_placement}} and as an artifact in Zenodo~\cite{tocze_2025_Artifact}. 

\section{Evaluation outcomes}
\label{sec:EvaluationOutcomes}
This section presents and discusses the outcomes of the evaluation presented in Section \ref{sec:Evaluation}. First, an initial study is presented in Section \ref{sec:results}. Then, the impact of load level heterogeneity and function instance availability are studied in Sections \ref{sec:influenceHeterogeneity} and \ref{sec:InfluenceAvailableInstances}. Next, the placement decisions obtained are discussed in Section \ref{sec:influenceCharacteristics}. Finally, a full co-location case and a generalization with regards to beginning device are presented in Section \ref{sec:specificAndgeneralization}.

\subsection{Initial study}

\label{sec:results}

In this section the initial group is considered. It means that both edge devices and communication links load levels are randomly chosen with a Normal distribution with a mean varying between 0 and 100 and a standard deviation of 10. As all groups, it contains 40 runs per combination of utilization level and optimization objective. 

\begin{figure}
     \centering         \includegraphics[width=\columnwidth]{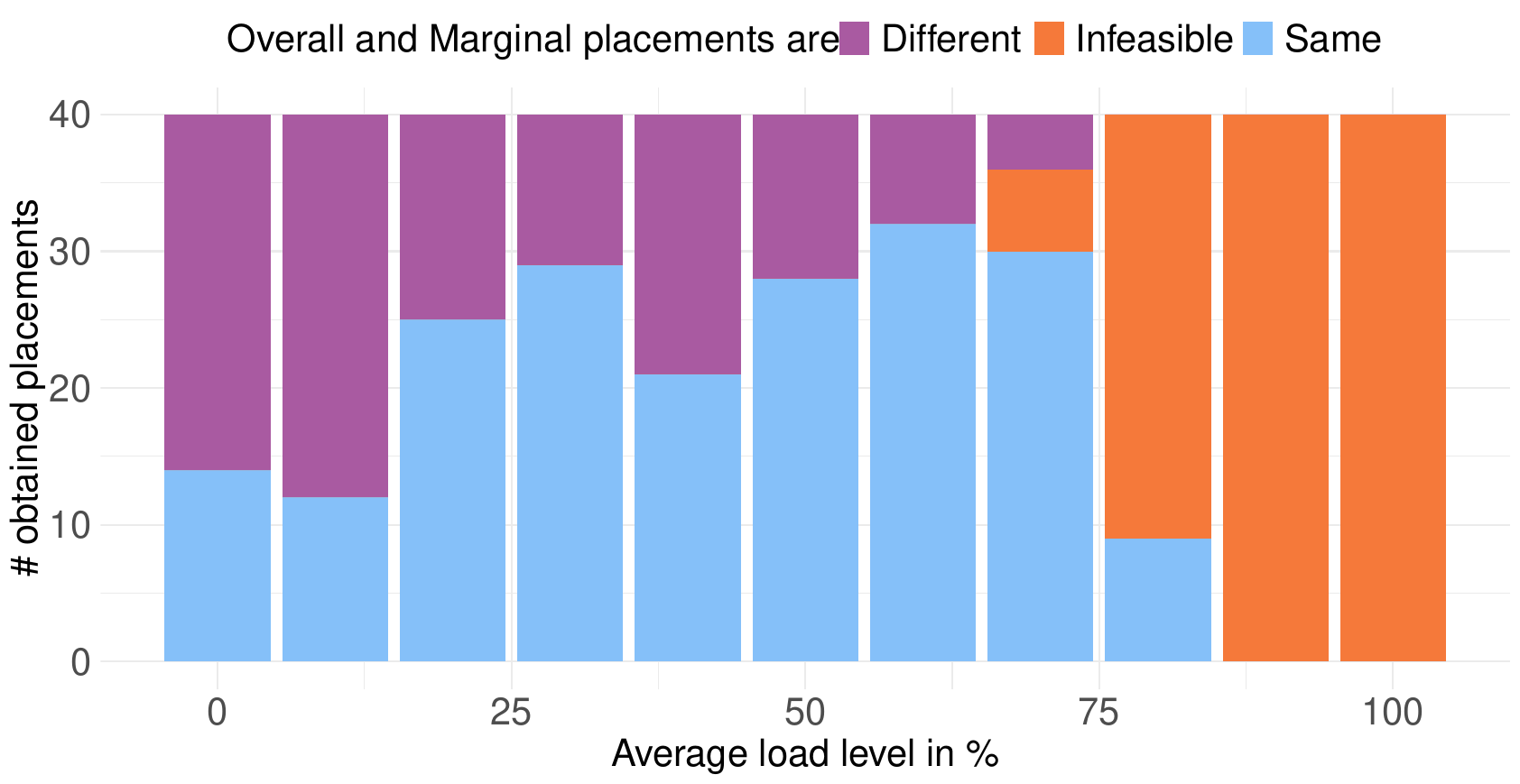}
         \caption{Categorization of the placements for a given request with varying load.}   
        \label{fig:initial_hist_different}
\end{figure}

The first question that we want to answer with these initial runs is: do the two optimization objectives actually lead to different placement decisions? Figure \ref{fig:initial_hist_different} shows the number of times per load level in which the placement is  infeasible/the same/different between the two optimization objectives. 

On this figure, it can be seen that it is indeed the case that optimizing for different objectives will lead to different placement decisions. How often this is the case varies strongly between the load levels  (from 0\% to 70\%). 

We now look into a specific run in order to analyze it in depth. The aim is to identify potential factors influencing the two 
objectives leading to different placement solutions. 

The specific run is randomly chosen among the 40 runs obtained for load level 50\% where the placement solution differs with respect to the two optimization objectives. This was the run with number 12. We select this load level as it represents a case when the edge infrastructure is neither underloaded nor overloaded. When the infrastructure is underloaded, there will be more devices that are idle, and hence, it is easier to find placements that differ between the two objectives (this is confirmed by Figure \ref{fig:initial_hist_different}). When the infrastructure is overloaded, the number of placements that still meet the deadline requirements is limited, hence it is less likely that the placement will be different between the two objectives. It is therefore interesting to look at load levels that are in between. 
Table \ref{tab:results} summarizes the results obtained for this specific run.

\begin{table}[]
    \centering
    \begin{tabular}{|l|c|c|}
    \hline
        Optimization objective & Overall energy & Marginal energy \\
        \hline
        Request completion time  & 92 ms      &  90 ms \\
        \hline
        Overall energy consumed  & 2240 J      & 2289 J  \\
        \hline
        Marginal energy consumed  & 906 J      & 850  J \\
        \hline
        Solver execution time & 220 ms   & 70 ms \\
        \hline
    \end{tabular}
    \caption{Evaluation results for one request placement.}
    \label{tab:results}
\end{table}

As seen in Table \ref{tab:results}, the different optimization objectives lead to similar request completion times. Regarding the energy metrics, the Overall alternative leads to 2.2\% less overall energy consumed  (- 49 J) than the Marginal one and the Marginal alternative leads to 6.6\% less marginal energy consumed (- 56 J) than the Overall one. To understand the actual decisions better, the placement solutions obtained are analyzed from the logs to understand how the difference appears. 
The placement solution differs in the following way between the different optimization objectives: functions 1, 3, and 4 are placed in the same function instances for both objectives, only the placement of function 2 is different. The difference in energy consumption shown in Table \ref{tab:results}, which is small but non-negligible, is therefore the result of only one of the four functions being placed in a different way.

In the logs, we look deeper at the placement, and especially the placement of F2. It is placed on device 5 in the Overall case, and on device 7 in the Marginal case. 
For the run considered, the devices 5 and 7 differ in the following ways:
\begin{itemize}
    \item Device 5 is less loaded than device 7. 
    \item The links used when choosing device 5 are less loaded than when using device 7. 
    \item The latency on the links used when choosing device 5 is higher than when using device 7. 
\end{itemize}

Therefore, the Overall energy metric seems to be able to leverage the time remaining to the deadline to pick edge devices located further away (leading to a higher completion time) but with lower utilization, and hence lower overall energy consumption. On the contrary, the Marginal energy metric favours a placement that is using edge devices with a higher utilization and closer.  

\begin{figure}
    \centering
    \includegraphics[width=0.9\columnwidth]{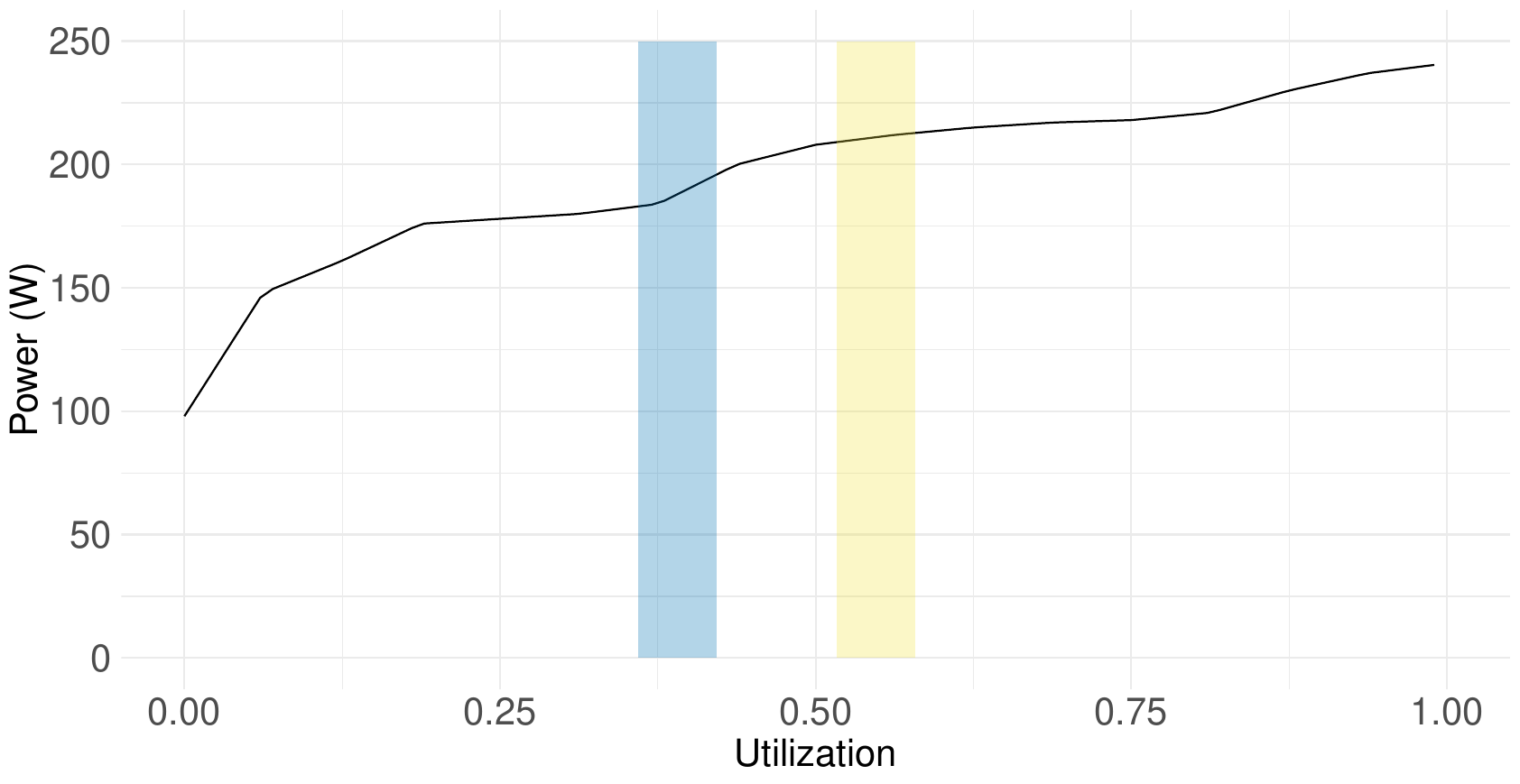}
    \caption{Power profile of a Parasilo device according to \cite{Ahvar_EstimatingEnergy}}.
    \label{fig:powerProfileParasilo}
\end{figure}

To understand the logic behind the Marginal placement, it is useful to look at the considered power profile for edge devices. It is shown in Figure \ref{fig:powerProfileParasilo}. The highlights correspond to the devices 5 (in blue) and 7 (in yellow), where the start of the highlight is the utilization before the function instance is placed and the end, the utilization after the placement. It can be seen that the increase in power is larger when placing on device 5 than on device 7, hence the choice of the Marginal alternative to place on device 7. 

These initial observations are inline with the envisioned strategies associated with the different metrics (see Section \ref{sec:objectiveFunction}), and give directions for further studies of factors influencing the different placement decisions. We present several of such studies in the coming sections. 
In this article, we focus on studies of factors related to the edge devices. 
Moreover, to limit the number of parameters varying and ease the analysis, we perform the following studies with a link utilization of 0\% on all links. 
As shown in Figure \ref{fig:initial_hist_different}, the load level is strongly influencing the placement decisions and whether they are different. Therefore, for the following studies, we show the results for varying load levels (i.e. different means of the Normal distribution from Section \ref{sec:loadScenario}) for edge devices. 

\subsection{Impact of heterogeneity} 
\label{sec:influenceHeterogeneity}

In this section, we first want to vary the heterogeneity of the load, i.e. whether the load is fixed at the same level for all edge devices (no heterogeneity) or very different, some edge devices with a low load and some having a high load (high heterogeneity). This corresponds to rows 2, 3 and 4 in Table \ref{tab:evaluation_groups}.

\begin{figure*}
     \centering
     \begin{subfigure}[b]{0.3\textwidth}
         \centering
         \includegraphics[width=\textwidth]{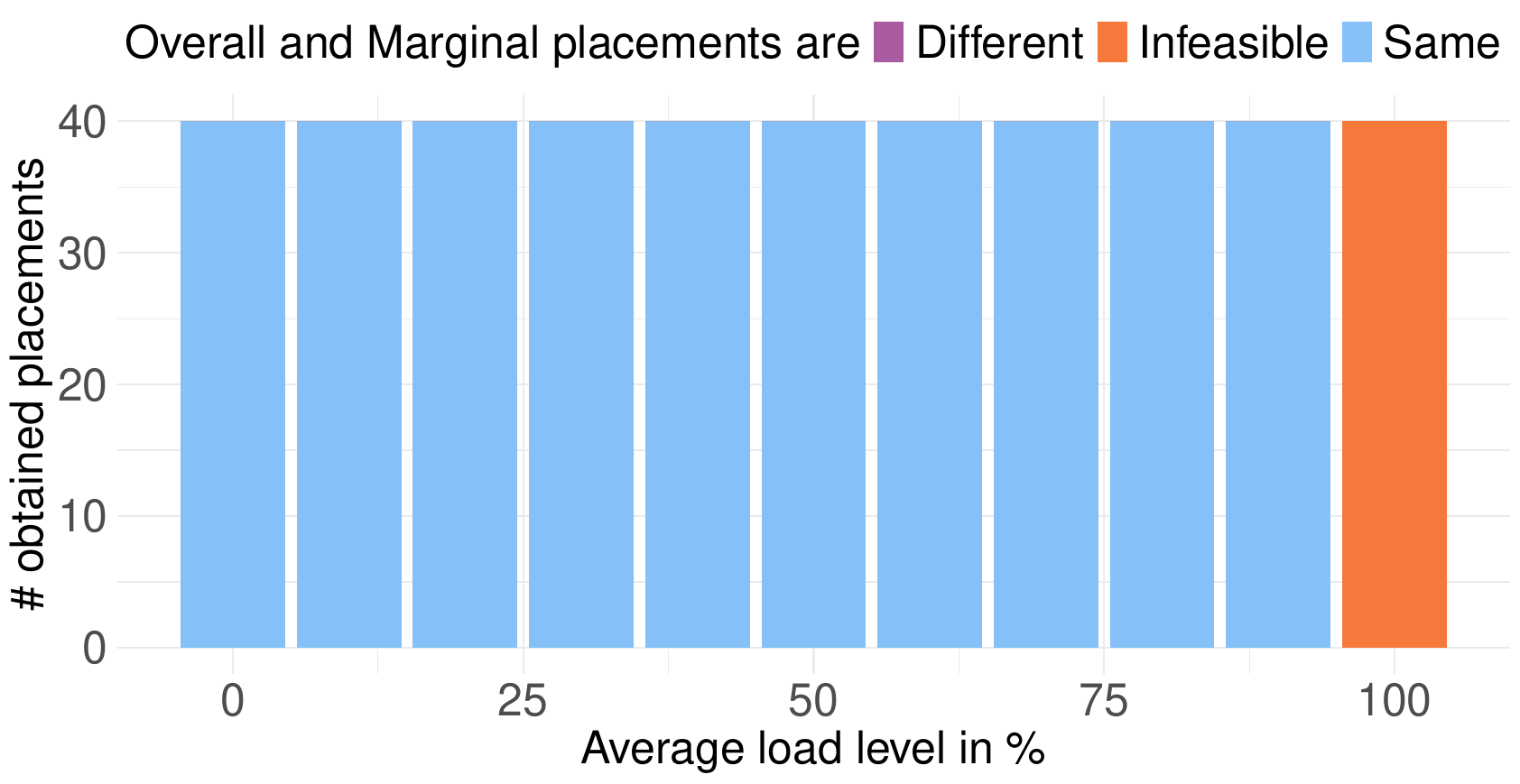}
         \caption{Fixed load at load level}
         \label{fig:loadLevelHomogeneous}
     \end{subfigure}
     \hfill
     \begin{subfigure}[b]{0.3\textwidth}
         \centering
         \includegraphics[width=\textwidth]{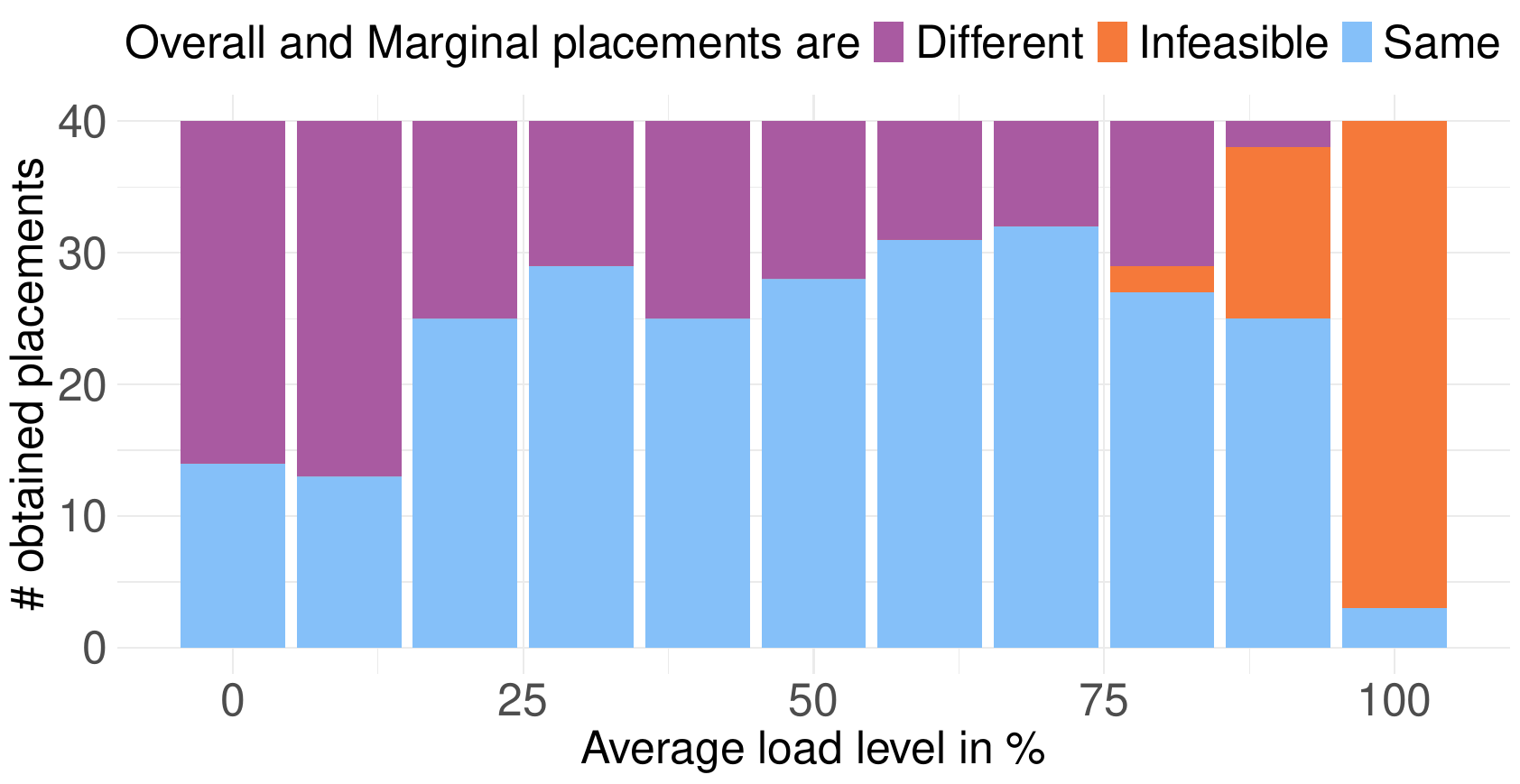}
         \caption{$\mathcal{N}$(load level, 10)}
         \label{fig:loadLevelHeterogeneous}
     \end{subfigure}
     \hfill
     \begin{subfigure}[b]{0.3\textwidth}
         \centering
         \includegraphics[width=\textwidth]{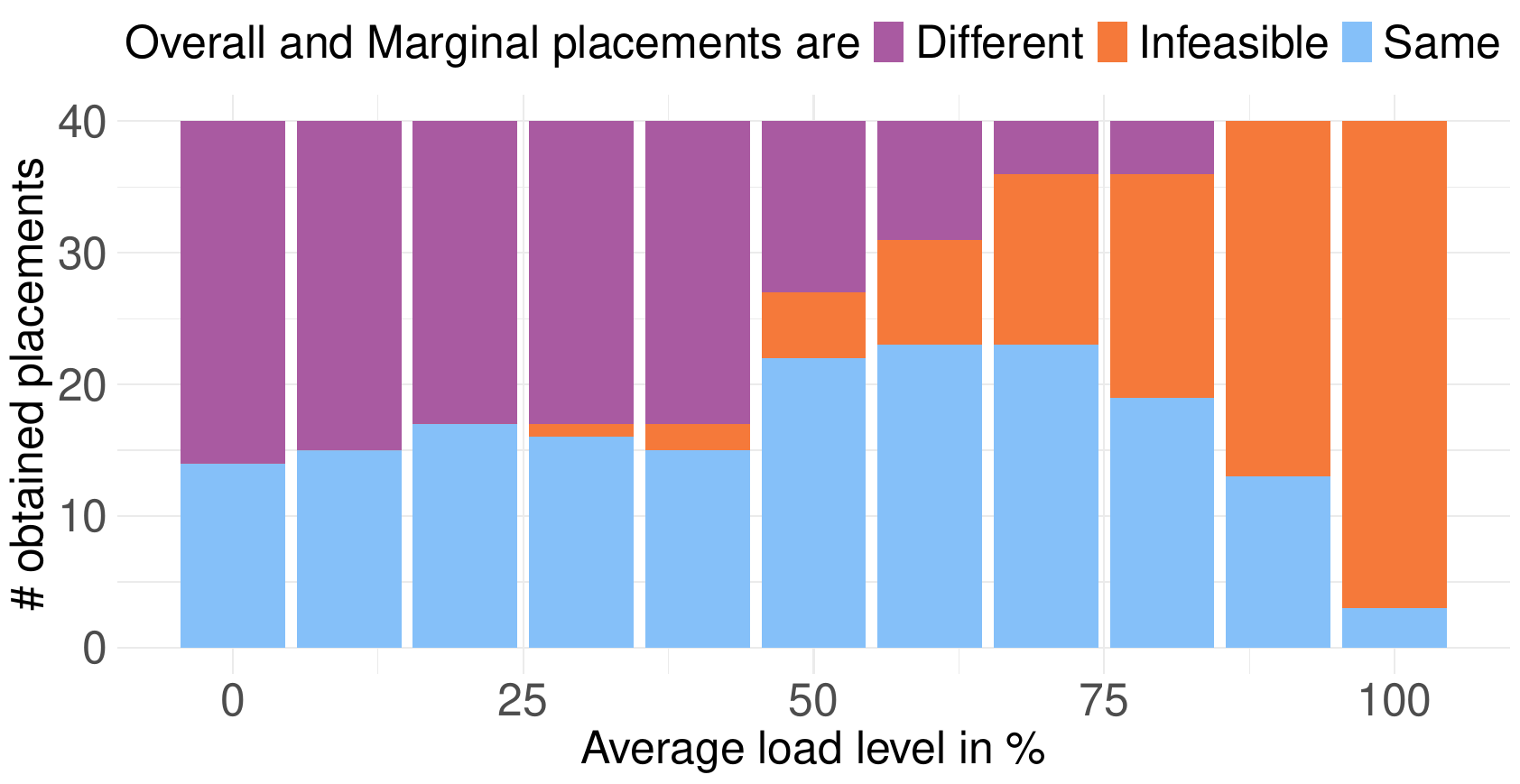}
         \caption{$\mathcal{N}$(load level, 30)}
         \label{fig:loadLevelHeterogeneous30}
     \end{subfigure}
        \caption{Categorization of the placements obtained for a given request with different edge device loads. }
        \label{fig:loadLevels}
\end{figure*}

The outcome after running 40 runs per load level and objective  is shown in Figure \ref{fig:loadLevels}, for three different load distributions. 
Each subfigure shows on the y-axis a categorization of the resulting 40 request placements in different load conditions. The load thus varies in two ways: first, the x-axis consists of different load levels, each bar representing one load level (0 to 100\% in steps of 10) for a group of runs. Second, the load level is used as an input to the distribution which introduces different loads per run across the 40 runs. 

In Figure \ref{fig:loadLevelHomogeneous}, the load is exactly the same on all edge devices, corresponding to the load level (this is row 3 in Table \ref{tab:evaluation_groups}). In that case, which objective is chosen does not matter, the placement will always be the same. Hence the absence of load heterogeneity leads to an absence of difference between the two optimization objectives. 

Figure \ref{fig:loadLevelHeterogeneous} shows the placement obtained when the edge device load varies between runs with a standard deviation of 10 (low heterogeneity). This is the same standard deviation that was used in the initial study. Here we see that introducing load heterogeneity between the devices does lead to different placement decisions. 

The difference between the edge device loads can be increased by increasing the standard deviation of the Normal distribution used from 10 to 30 (Figure \ref{fig:loadLevelHeterogeneous30}). This corresponds to a case where the available resources differ to a greater extent between the edge devices. In this case, the two optimization objective alternatives overall lead to different placements to a greater extent.
It can be noted that it also means a greater extent of infeasible placements for the higher load levels. 
As an edge infrastructure in practice will exhibit heterogeneity, this emphasizes the importance of carefully choosing the energy metric to optimize against. 

\subsection{Impact of availability of function instances}
\label{sec:InfluenceAvailableInstances}

The next aspect studied is the availability of the function instances, i.e. the number of instances that are available for each function when a service request arrives. A higher number of function instances means more possible choices for the placement of the request and a lower one restricts the options of the optimization. 

We therefore reproduce the experiment presented on row 2 of Table \ref{tab:evaluation_groups}, which had 2 available function instances, with 4 available function instances (row 5) and 6 available function instances (row 6). 
The function instances are placed in a pre-defined way and so that two new instances of all the functions in the chain are added for each new experiment, but the already placed ones are not changed. 

\begin{figure*}
     \centering
     \begin{subfigure}[b]{0.3\textwidth}
         \centering
         \includegraphics[width=\textwidth]{images/40rep_DeviceOnlyVar_2replicas_Std10_RequestDevice4_v2.pdf}
         \caption{2 avail. function instances per function}
         \label{fig:functionInstance2}
     \end{subfigure}
     \hfill
     \begin{subfigure}[b]{0.3\textwidth}
         \centering
         \includegraphics[width=\textwidth]{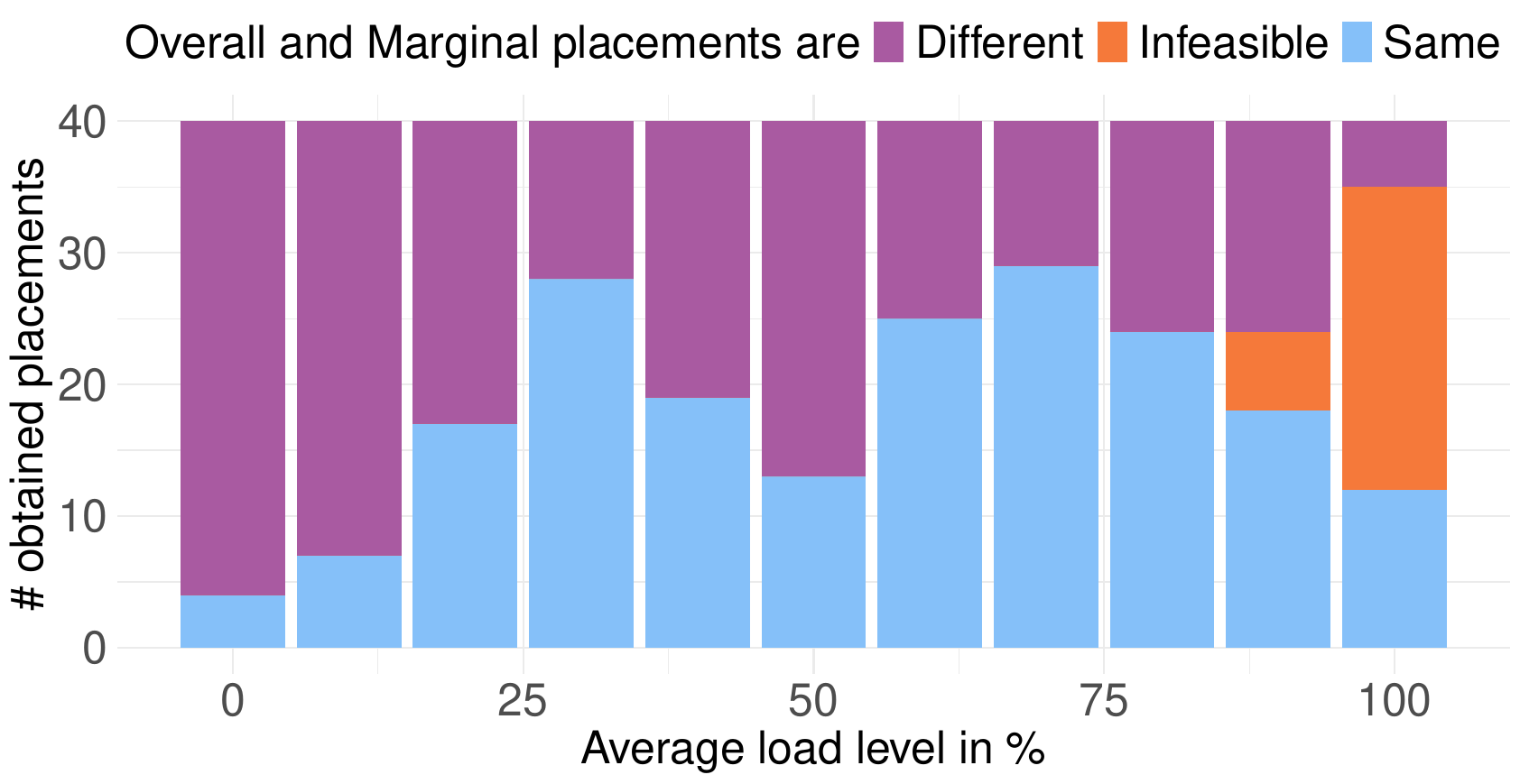}
         \caption{4 avail. function instances per function}
         \label{fig:functionInstance4}
     \end{subfigure}
     \hfill
     \begin{subfigure}[b]{0.3\textwidth}
         \centering
         \includegraphics[width=\textwidth]{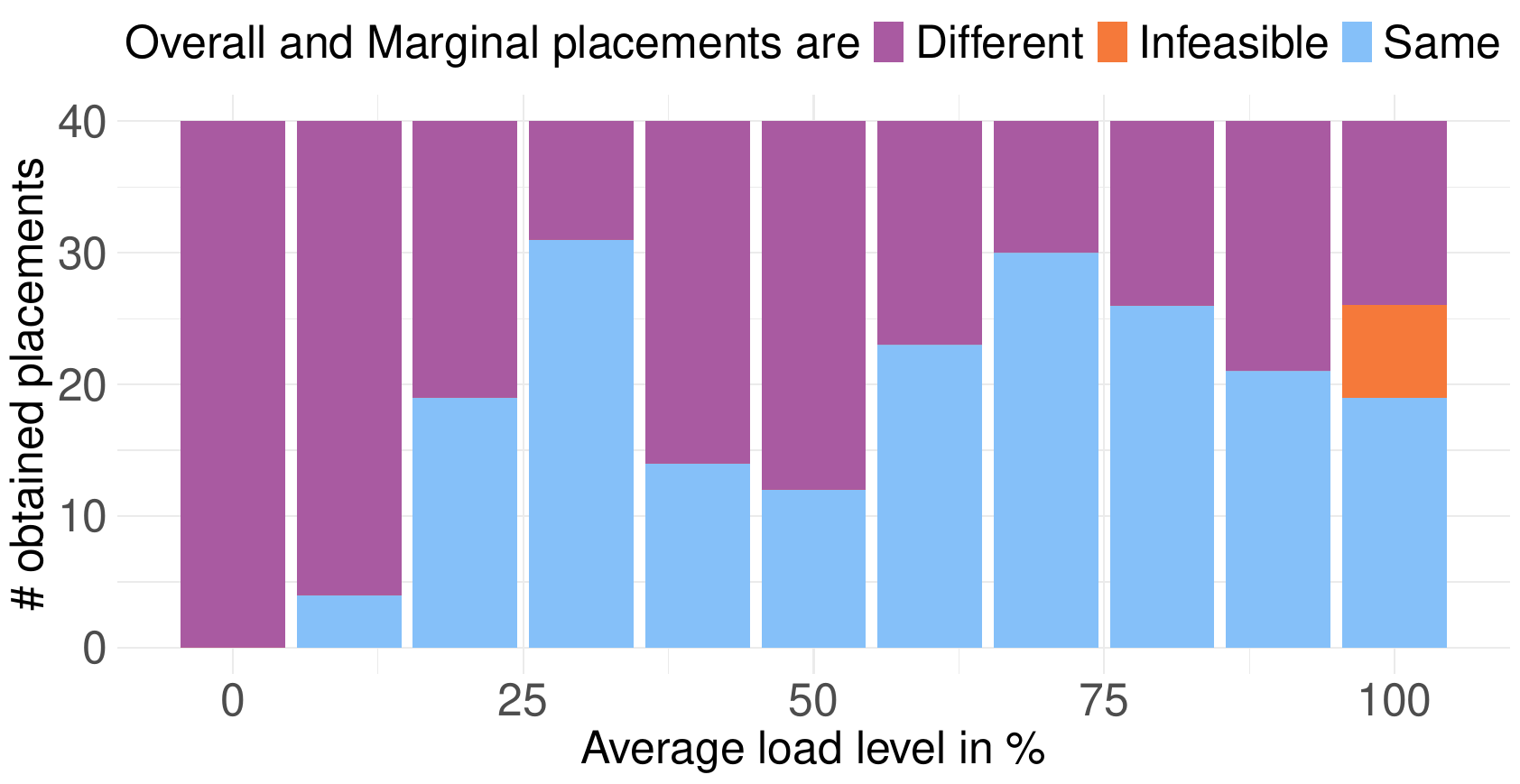}
         \caption{6 avail. function instances per function}
         \label{fig:functionInstance6}
     \end{subfigure}
        \caption{Categorization of the placements obtained for a given request with varying load and function instance availability.}
        \label{fig:functionInstances}
\end{figure*}

The outcomes are depicted in Figure \ref{fig:functionInstances}. As the number of function instances increases, from 2 to 6 instances per function (Figures \ref{fig:functionInstance2} to \ref{fig:functionInstance6}), the number of cases in which the different optimization objective alternatives lead to different placements overall increases, especially between 2 and 4 function instances. The higher the number of function instances, the less infeasible placements.

We look deeper into the difference between the placements and look at how large the resulting relative difference in energy consumption is.
Table \ref{tab:DifferenceExtent}  shows the 10th and 90th percentile values of the relative difference, i.e. the increase in percentage compared to the optimal value for each energy metric. For a given availaibility group (each line), the relative difference is in the same range for all load levels, apart from a few outliers. Violin plots showing the distribution of the difference can be found in Appendix \ref{ap:extraCurves_reldiff}. Table \ref{tab:DifferenceExtent} also shows that the range of the difference increases with the size of the solution space (more function instances available). 

\begin{table}[]
    \centering
    \begin{tabular}{|c|c|c|c|c|}
    \hline
        \textbf{Availability} & \multicolumn{2}{c|}{\textbf{Overall energy}}& \multicolumn{2}{c|}{\textbf{Marginal energy}}  \\
        \hline
        \multirow{2}{*}{2 function instances} & 10th & 0.3\% &10th  & 0.6\% \\
        \cline{2-5}
        & 90th  & 6.4\% &90th  & 56.9\% \\
        \hline
\multirow{2}{*}{4 function instances} & 10th  & 0.1\% &10th  & 0.5\% \\
        \cline{2-5}
        & 90th  & 12.0\% &90th  & 65.0\% \\
        \hline
        \multirow{2}{*}{6 function instances} & 10th  & 0.3\% &10th  & 1.9\% \\
        \cline{2-5}
        & 90th  & 27.6\% &90th & 106.8\% \\
        \hline
        
    \end{tabular}
    \caption{Percentiles for the relative difference in energy consumption, when placements are different, all load levels.}
    \label{tab:DifferenceExtent}
\end{table}

In a nutshell, as the number of function instances increases, the number of load scenarios in which the different optimization objective alternatives lead to different placements increases. This effect is already significant for small solution spaces as shown in the evaluation. This suggests that it is relevant to carefully pick the energy metric used for optimization as the choice will have a higher influence on placement decision as the solution space widens. 

\subsection{Analysis of placement decisions}
\label{sec:influenceCharacteristics}

In Section \ref{sec:results}, we observed how the different optimization objectives led to different ways of placing the request. In particular, the Overall alternative seemed to favor devices located further away with lower load, while the Marginal alternative seemed to favor closer devices with current higher load, which is inline with the idea behind the different energy metrics. 
This flexibility of choosing different devices is possible as long as the completion time stays below the  deadline. 
Do these findings hold as well for the other evaluated scenarios? In this section, we study the resulting placements from rows 4, 5, and 6 of Table \ref{tab:evaluation_groups} to see whether the envisioned placement strategies are actually reflected in the placement decisions. 

\begin{figure*}
     \centering
     \begin{subfigure}[b]{0.3\textwidth}
         \centering
         \includegraphics[width=\textwidth]{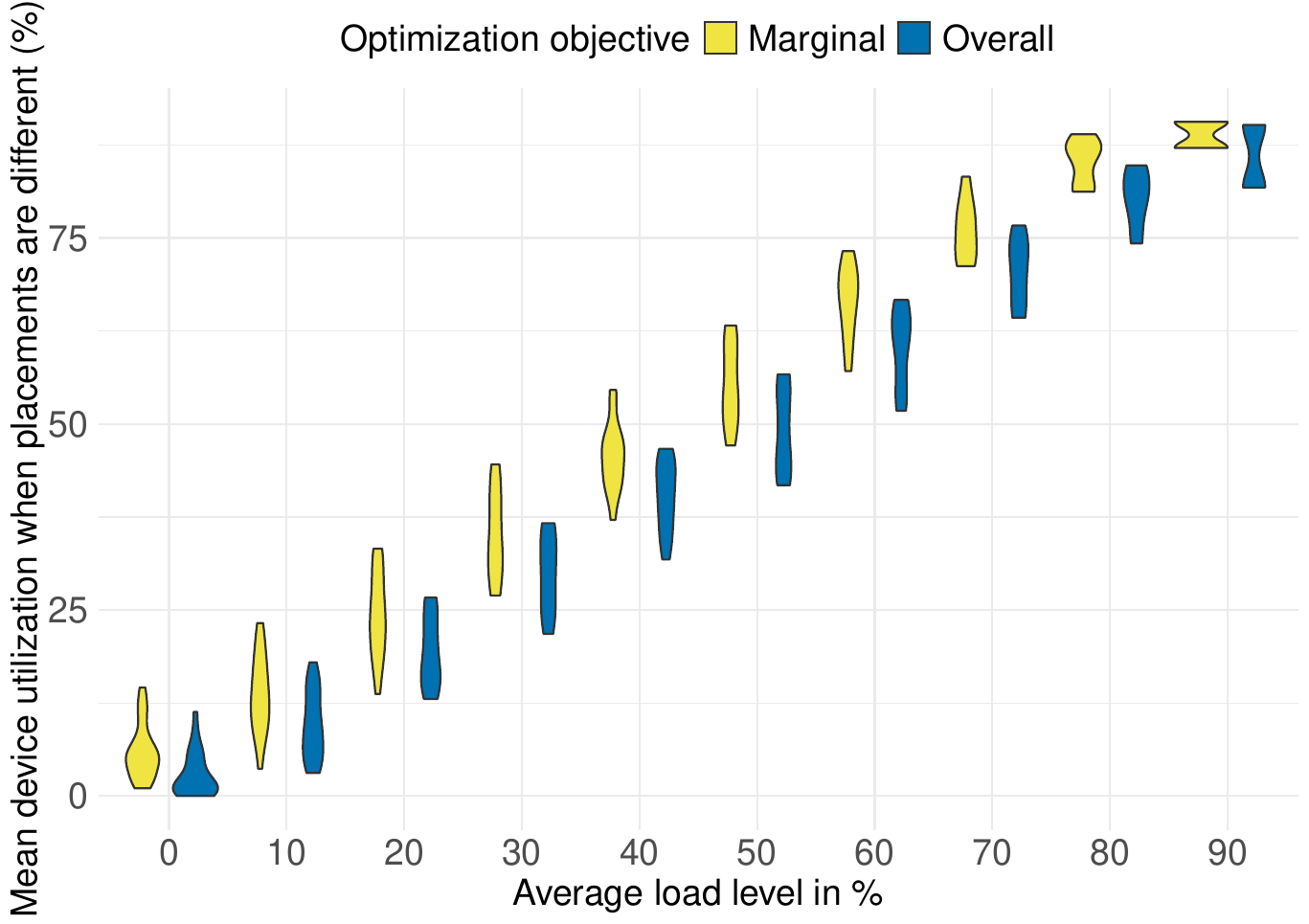}
         \caption{2 avail. function instances per function}
         \label{fig:utilization2}
     \end{subfigure}
     \hfill
     \begin{subfigure}[b]{0.3\textwidth}
         \centering
         \includegraphics[width=\textwidth]{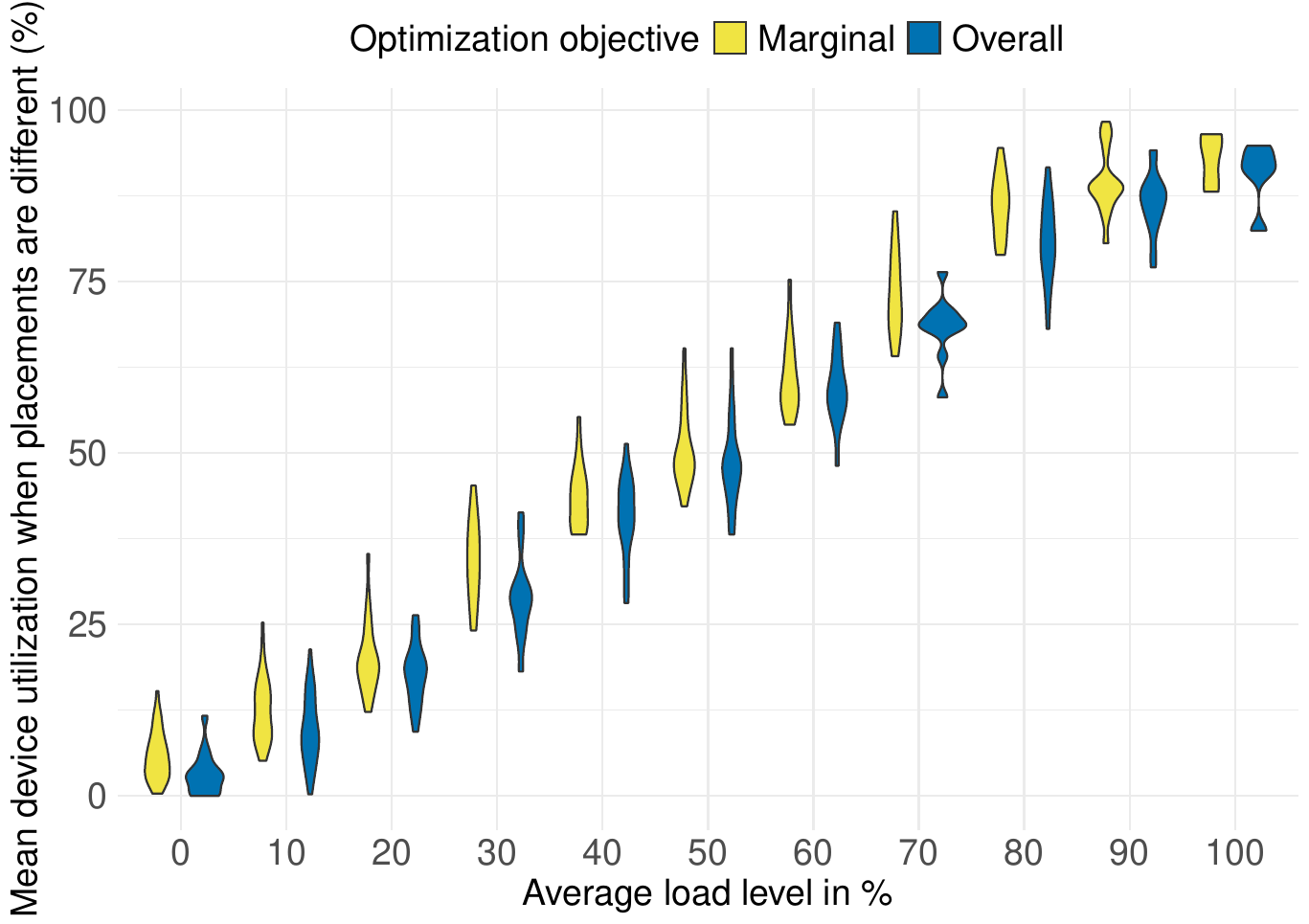}
         \caption{4 avail. function instances per function}
         \label{fig:utilization4}
     \end{subfigure}
     \hfill
     \begin{subfigure}[b]{0.3\textwidth}
         \centering
         \includegraphics[width=\textwidth]{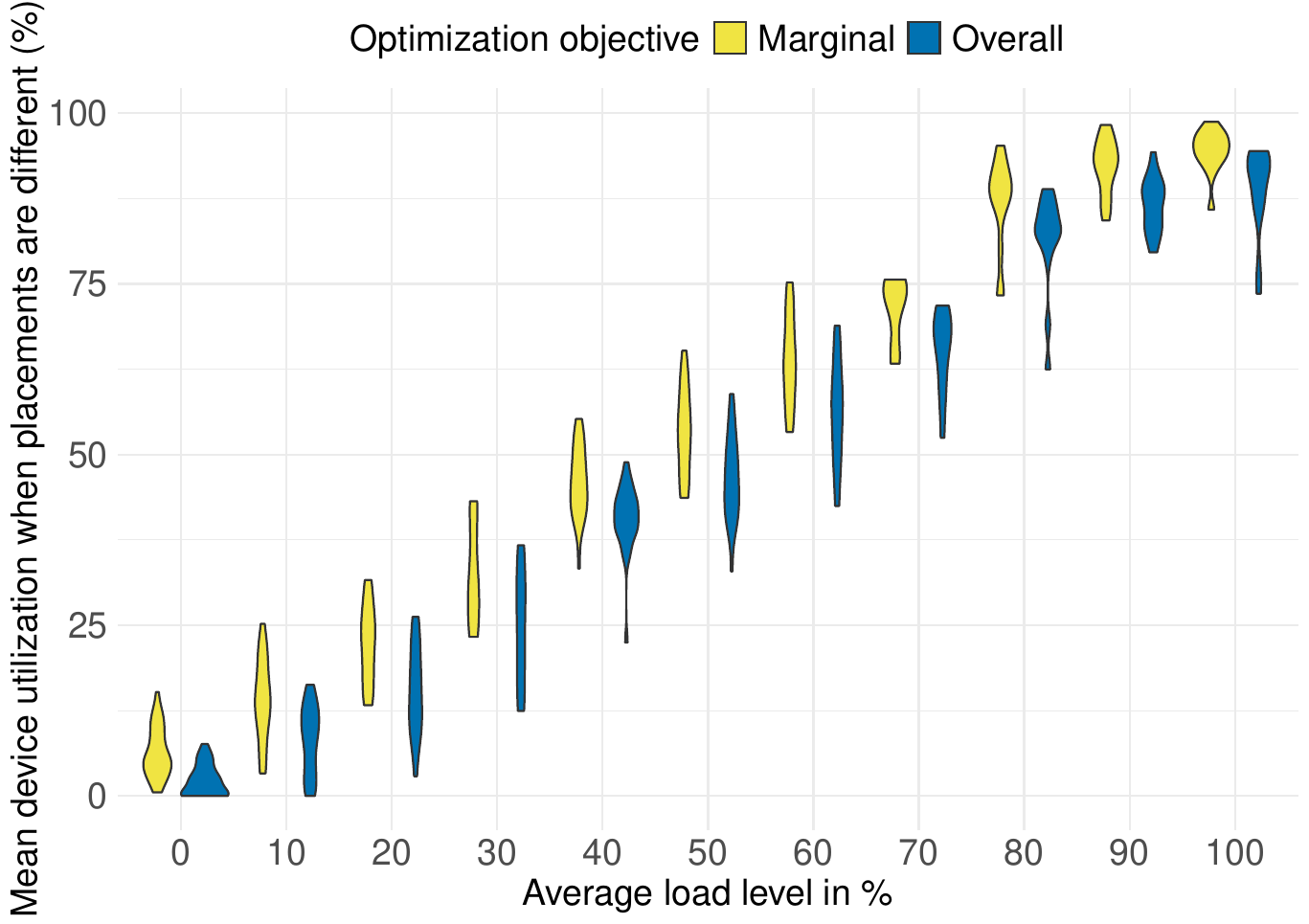}
         \caption{6 avail. function instances per function}
         \label{fig:utilization6}
     \end{subfigure}
        \caption{Average included device utilization  with varying function instance availability and infrastructure load levels.}
        \label{fig:deviceUtilization}
\end{figure*}

First, we look at the utilization of the chosen devices, focusing only on when the placements obtained are different. Figure \ref{fig:deviceUtilization} shows the distribution of the average utilization level of the chosen devices before the function instances are placed, using a violin plot. It shows that indeed, the chosen devices for the Marginal alternative usually have a higher utilization than those chosen for the Overall alternative. 

Next, we look at the request completion time to see whether one optimization objective is associated with a longer completion time, as a result of favoring devices located further away. For space reasons, the corresponding curves are shown in Appendix \ref{ap:extraCurbes_completion}. They show that except for higher load levels (>80\%), the completion times are usually close between the different alternatives. The Marginal alternative achieves on average a lower request completion time, except for load level zero. This suggests that Marginal may favor closer devices but as the difference is small, this needs to be investigated further. Another observation is that completion times decrease when more function instances are available.

In a nutshell, the evaluation outcomes suggest that optimization against the two considered metrics leads to the placement decisions that were envisioned for the configurations tested. Regarding the empirical observation that the Marginal alternative seem to favor closer devices, it has to be further investigated whether this is also the case in general. 

\subsection{Specific case and generalization}
\label{sec:specificAndgeneralization}
\subsubsection{Full co-location is possible}
\label{sec:FullColocation}

\begin{figure}
    \centering
    \includegraphics[width=\columnwidth]{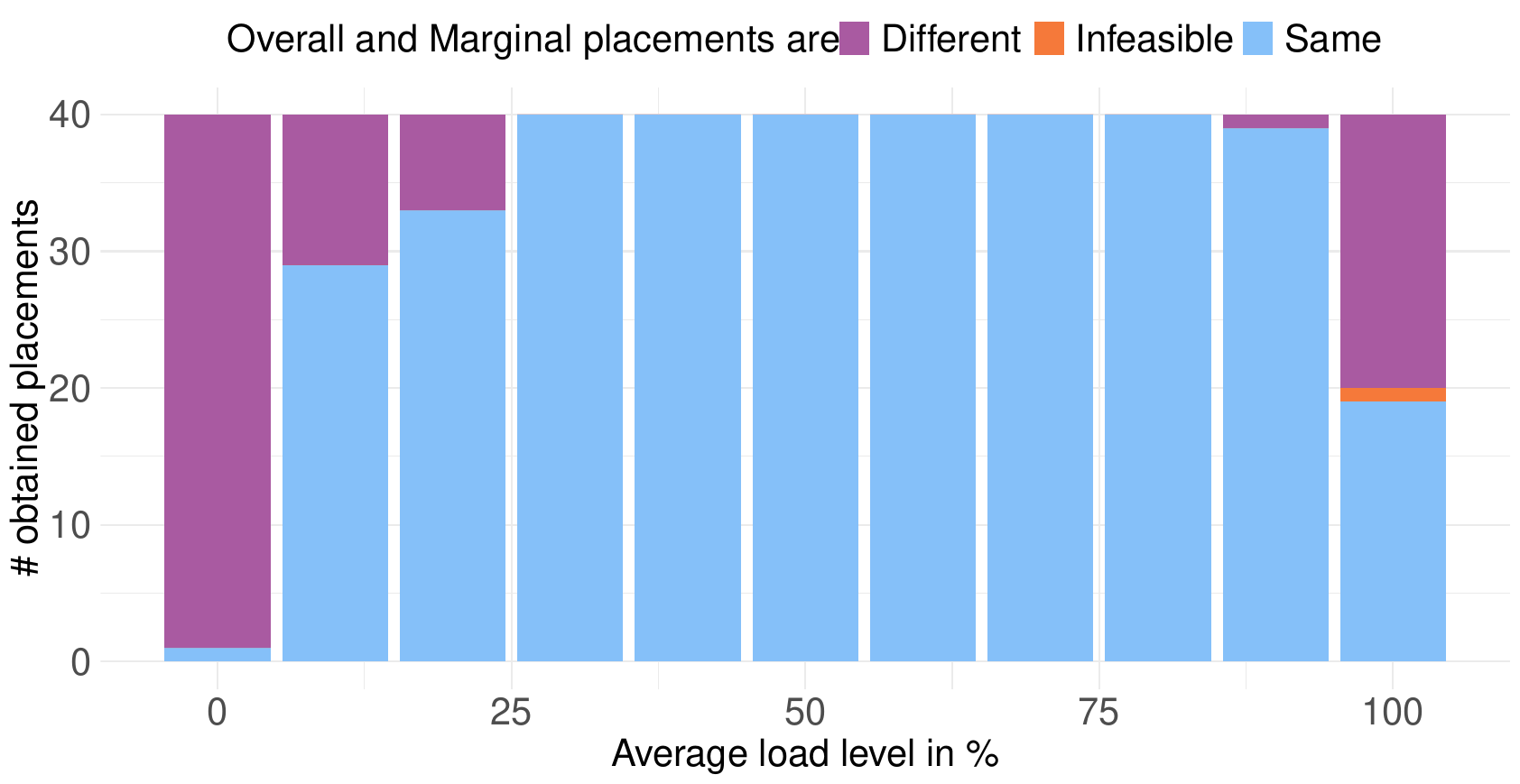}
    \caption{Categorization of the placements obtained for a given request where full co-location is possible.}
    \label{fig:FullColocation}
\end{figure}

One interesting specific case to study is when it is possible to execute all the functions on the beginning device (i.e. to fully co-locate the execution on the device receiving the request). We create such a scenario (row 7 of Table \ref{tab:evaluation_groups}) by changing the beginning edge device from row 6 and show the results in Figure \ref{fig:FullColocation}. Compared to Figure \ref{fig:functionInstance6}, the number of placements that are the same is a lot higher, which corresponds to when the beginning edge device is chosen for execution for both optimization objectives. The placements which are different at lower load levels corresponds to those runs where the utilization level of the beginning edge device is 0. In this case, the Marginal alternative will execute (possibly part of) the functions on other devices to avoid starting using the beginning device for computation. From a placement strategy perspective, one has to decide whether this is a wanted behavior as offloading to other devices has a cost in terms of networking, which also translates into energy consumption. 

\subsubsection{Random beginning device}
\label{sec:randomRequesting}

As the combination of beginning device and function instance placement can influence the results as shown in Section \ref{sec:FullColocation}, we investigate further whether the results presented in Sections \ref{sec:influenceHeterogeneity}, \ref{sec:InfluenceAvailableInstances}, and \ref{sec:influenceCharacteristics} are representative. Therefore, we conduct a new group of runs (rows 8 and 9) where the beginning edge device is selected randomly between repetitions of running all load levels for the two optimization objectives.   

This study confirms the previous findings, i.e. that higher load heterogeneity and number of available function instances lead to more different placements. Also it confirms the ability of the Marginal alternative to select devices with a higher utilization. Two interesting things can be noted: first, the number of infeasible placements increases for lower load levels, when only two function instances are available. This effect disappears when more function instances are available as the additional location covered allow the service to complete before the deadline no matter where the request comes in. Second, the completion times present a larger span and shows that both alternatives leverage the remaining time to the deadline to find the best placement for their objective. None of them is favoring closer devices over devices located further away.  
Due to space constraints, the corresponding figures are 
in Appendix \ref{ap:RandomRequestingDevice}.

\subsection{Discussion}
\label{sec:Discussion}

The goal of this work was to investigate whether using different energy metrics as optimization objectives could lead to different energy-saving microservice request placement decisions, according to different strategies. To perform this study, we formulate the problem as an ILP and solve it optimally for a small-scale problem. This method is used for several reasons. First, the choice of a small-scale edge infrastructure and microservice deployment  serves several purposes: 1) it is easier to analyze the outcomes as the solution space is limited, 2) it demonstrates the relevance of the idea in small setups while suggesting that it is even more useful for larger setups, and 3) it allows to solve the problem optimally in a reasonable time. Solving the problem optimally allows the conclusion to be drawn on the actual optimal and not some approximation. It is therefore a useful tool for understanding the proposed approach.  

Of course, to implement this approach of using energy metrics to achieve a given placement strategy in practice, solving the ILP using a solver is not feasible as it takes too long to execute. Even if the solver execution time for performing the evaluation presented was short (mostly in the range of hundred milliseconds, up to a few seconds), it is still too important considering that the service considered has a deadline at 100 ms. Since the problem is formulated as a TPP instance, it is possible to use the numerous very good (meta-) heuristics which have been developed for such problems. Example of these are tabu search or Ant Colony Optimization. The model presented in this work can then be used a a reference to know how close to optimality these heuristics perform.    

\section{Related works}
\label{sec:RelatedWorks}

The microservice architecture~\cite{Cerny_MicroserviceArchitecture} is being extensively considered in the neighbouring field of cloud computing. It is a cornerstone of both the cloud-native~\cite{CloudNativeFoundation} and serverless~\cite{Li_ServerlessComputing_Survey} computing. 
Several surveys have presented the microservice works in cloud computing, including focusing on anomaly detection~\cite{Soldani_AnomalyDetection_Survey} or  practical dimensions such as Kubernetes scheduling~\cite{Carrion_KubernetesScheduling_Survey}. 
Microservices are also increasily used in the edge computing paradigm. The fine-grained modularity they offer is for example an asset for IoT applications~\cite{Pallewatta_PlacementMicroservicesIoT_Survey}. 

Microservice request placement is tackled by Shameli-Sendi et al. ~\cite{Shameli-Sendi_1_OptimalPlacement,Shameli-Sendi_EfficientProvisioning}. They also use a TPP formulation for simultaneously placing multiple service requests in a cloud environment. Contrary to this work, they do not consider energy, nor that the request placement has to meet a deadline. They also assume that all devices are able to provide all function instances, which is not the case in edge and therefore in our work. 

Schneider et al~\cite{Schneider_DistributedOnline} study placement problems at the edge and propose a solution for service coordination, which they define as combining the online scaling, placement, scheduling and routing of a service request. Their solution uses distributed deep reinforcement learning. They do not consider energy and assume that you can add function instances to serve the request. 

Wang et al.~\cite{Wang_MicroserviceOriented} study the placement of function instances and 
requests in the context of the Internet of Vehicles. They present a two-layer system including three algorithms to 1) place the service requests to devices, 2) enable different service requests to use the same function instance and 3) reduce the number of service instances when they are not needed anymore. Instead, we focus on the first layer to understand how optimizing towards energy instead of performance can be used for different placement strategies. 

Russo Russo et al.~\cite{RussoRusso_QoSAwareOffloading} propose a microservice request placement also with two levels. The first level is a simple heuristic algorithm that places every incoming request according to a probability distribution. The second level is optimizing these probabilities at regular time intervals. The optimization is performed using linear programming considering monetary cost and resource availability constraints. The aim is to maximize the requests satisfying their response time requirements.  Contrary to our work, the energy consumption is not considered. Moreover, the optimization has a different purpose as it is not directly performing the placement.

There are several works considering energy consumption within edge computing, e.g. \cite{Gnibga_LatencyEnergyCarbonAware,Hadjur_Beekeeping,Li_SustainableEdgeRenewable,Perin_SustainableEdgeComputing}. To evaluate the proposed methods and techniques, relevant energy models are required. 
Baccarelli et al.~\cite{Baccarelli_EcoMobiFog} propose models for the energy consumption of both devices and communication links. Their focus is on virtualized and multi-core devices in the context of 5G. 
Similarly, Ahvar et al.~\cite{Ahvar_EstimatingEnergy} propose energy models for different types of so-called cloud-related architectures. It also includes models for both the devices and the communication links. Their aim is to compare different types of architectures (more or less distributed) with regards to their energy consumption. 
To the best of our knowledge, no study considers energy-centric microservice request placement.

\section{Conclusion}
\label{sec:Conclusion}

In this work, we propose and investigate an energy-centric approach to place microservice requests: optimizing for  different energy metrics. We formulate the problem as an instance of the TPP problem and solve the corresponding ILP as a tool to study the effect of different metrics and variations in the problem space. 

Our evaluation suggests that optimizing using the  two proposed  energy metrics does lead to the envisioned placement strategies. In both cases, the completion time is a leverage to select devices that are most suitable for the objective but located further away. Higher heterogeneity in the system load and number of available function instances are factors leading to different placement decisions between the two objectives.

While our work initiates the study of energy metrics for energy-centric placement strategies, more work remains in the area. 
For example, a two-tier placement strategy combining the two optimization objectives based on the current system state would be interesting to develop. 
Moreover, alternative fast decision strategies for finding a good enough solution should be looked into to implement the approach in practice. 

\begin{acks}
This work is supported by the Swedish national graduate school in computer science (CUGS). The second author was supported by ELLIIT, Excellence Center at Linköping-Lund on Information Technology.
\end{acks}

\bibliographystyle{ACM-Reference-Format}
\bibliography{sample-base}

\appendix

\section{Additional outcome visualizations}

\subsection{Distribution of the relative difference}
\label{ap:extraCurves_reldiff}

\begin{figure*}
     \centering
     \begin{subfigure}[b]{0.3\textwidth}
         \centering
         \includegraphics[width=0.9\textwidth]{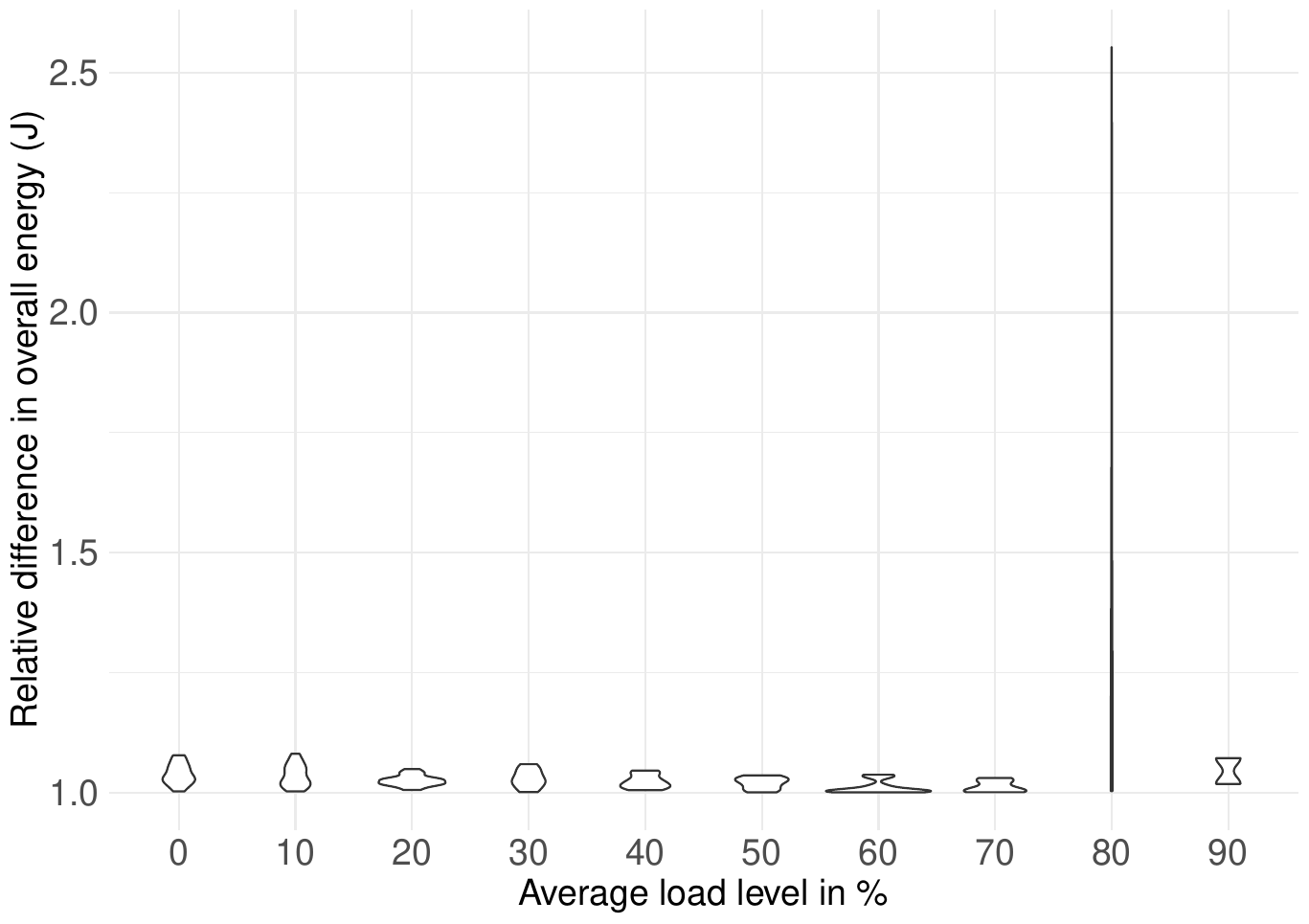}
         \caption{2 avail. function instances per function}
         \label{fig:functionInstance2RelativeDifference}
     \end{subfigure}
     \hfill
     \begin{subfigure}[b]{0.3\textwidth}
         \centering
         \includegraphics[width=0.9\textwidth]{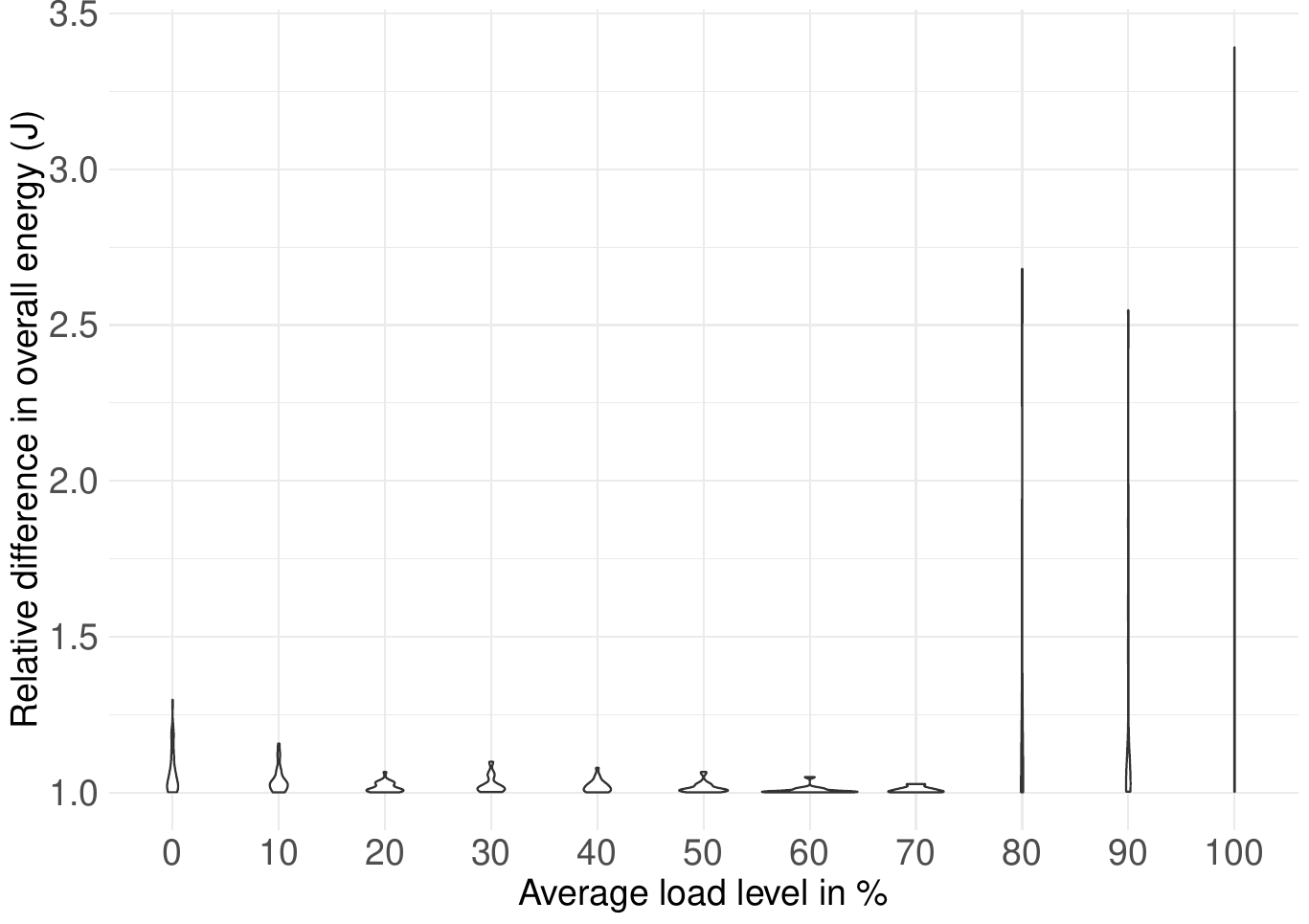}
         \caption{4 avail. function instances per function}
         \label{fig:functionInstance4RelativeDifference}
     \end{subfigure}
     \hfill
     \begin{subfigure}[b]{0.3\textwidth}
         \centering
         \includegraphics[width=0.9\textwidth]{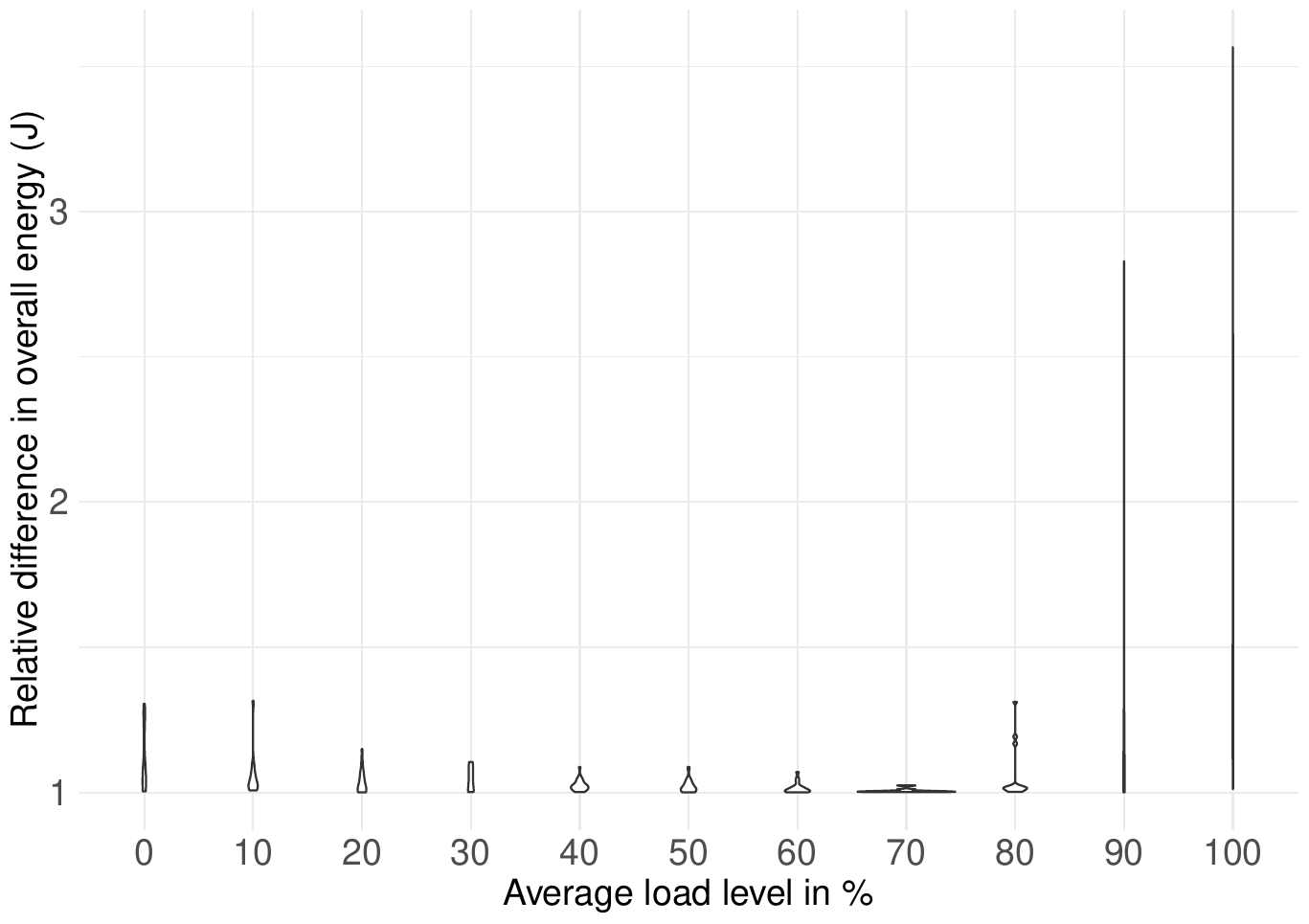}
         \caption{6 avail. function instances per function}
         \label{fig:functionInstance6RelativeDifference}
     \end{subfigure}
        \caption{Overall energy relative difference for a given request with varying load and function instance availability (when the resulting placement is different).}
        \label{fig:functionInstancesRelativeDifference}
\end{figure*}

\begin{figure*}
     \centering
     \begin{subfigure}[b]{0.3\textwidth}
         \centering
         \includegraphics[width=0.9\textwidth]{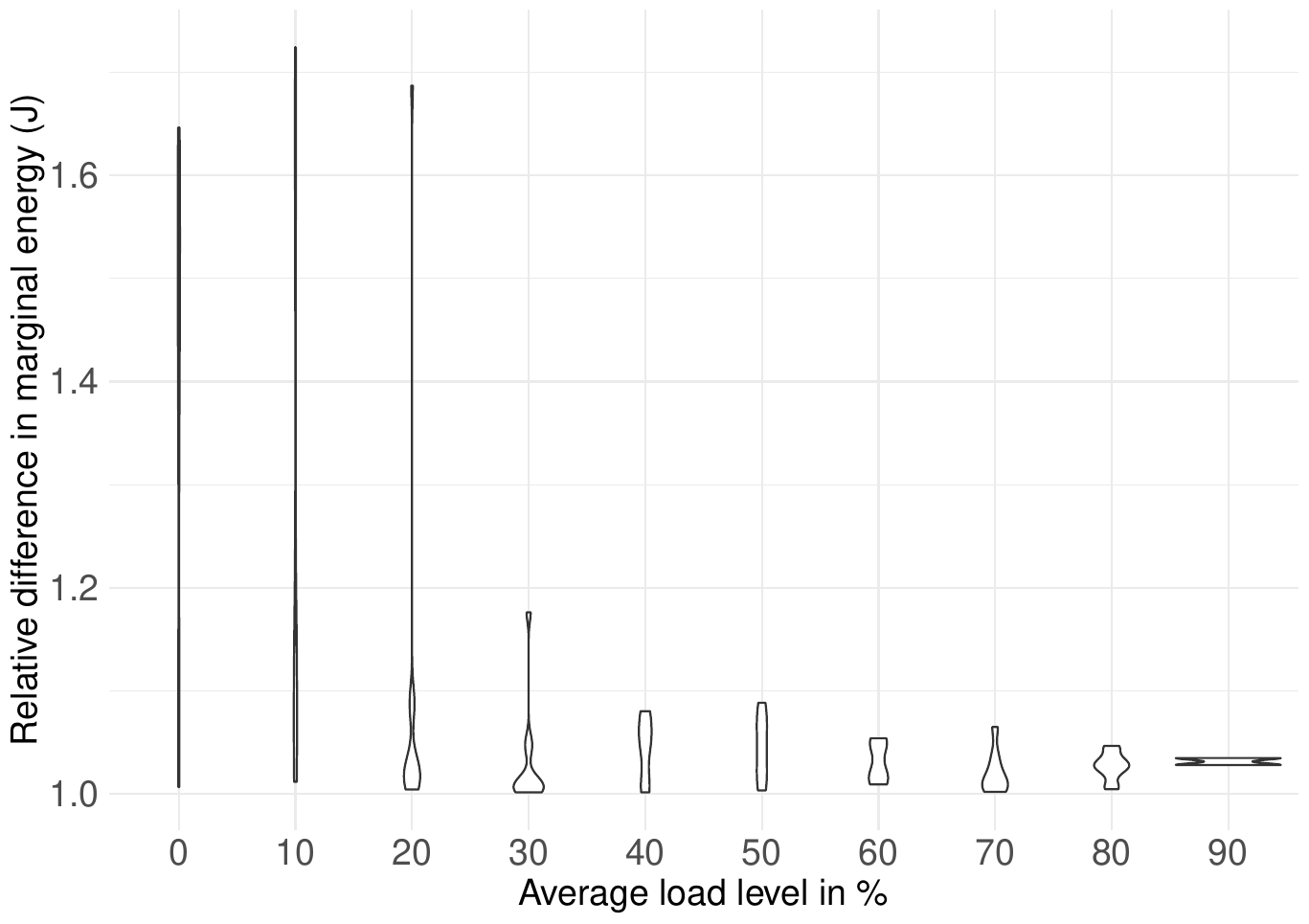}
         \caption{2 avail. function instances per function}
         \label{fig:functionInstance2RelativeDifferenceMarg}
     \end{subfigure}
     \hfill
     \begin{subfigure}[b]{0.3\textwidth}
         \centering
         \includegraphics[width=0.9\textwidth]{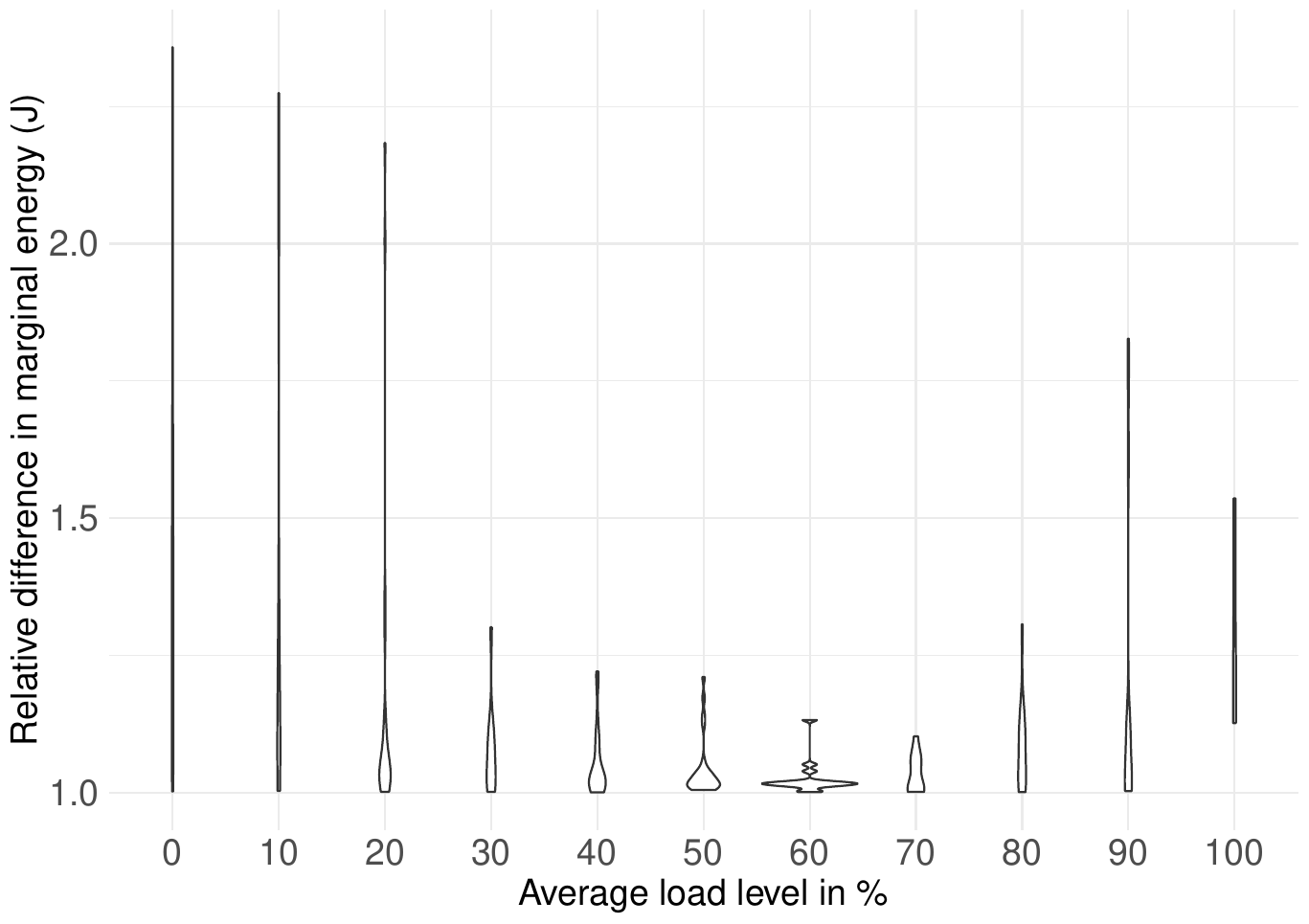}
         \caption{4 avail. function instances per function}
         \label{fig:functionInstance4RelativeDifferenceMarg}
     \end{subfigure}
     \hfill
     \begin{subfigure}[b]{0.3\textwidth}
         \centering
         \includegraphics[width=0.9\textwidth]{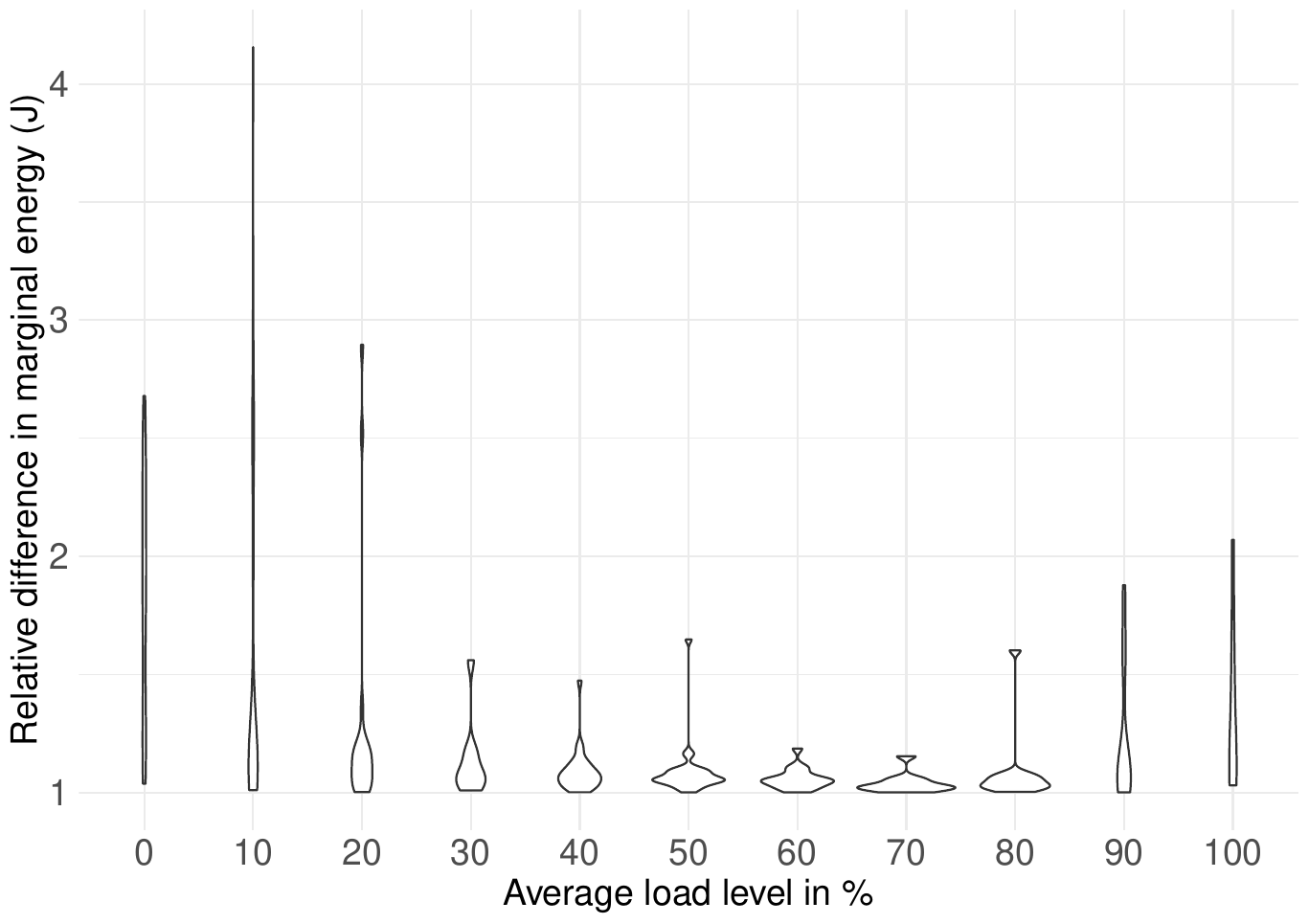}
         \caption{6 avail. function instances per function}
         \label{fig:functionInstance6RelativeDifferenceMarg}
     \end{subfigure}
        \caption{Marginal energy relative difference for a given request with varying load and function instance availability (when the resulting placement is different).}
        \label{fig:functionInstancesRelativeDifferenceMarg}
\end{figure*}

Figure \ref{fig:functionInstancesRelativeDifference} shows the relative difference when placement are different in terms of overall energy. Figure \ref{fig:functionInstancesRelativeDifferenceMarg} shows the relative difference when placement are different in terms of marginal energy. 

\subsection{Distribution of the completion time}
\label{ap:extraCurbes_completion}

\begin{figure*}
     \centering
     \begin{subfigure}[b]{0.3\textwidth}
         \centering
         \includegraphics[width=0.9\textwidth]{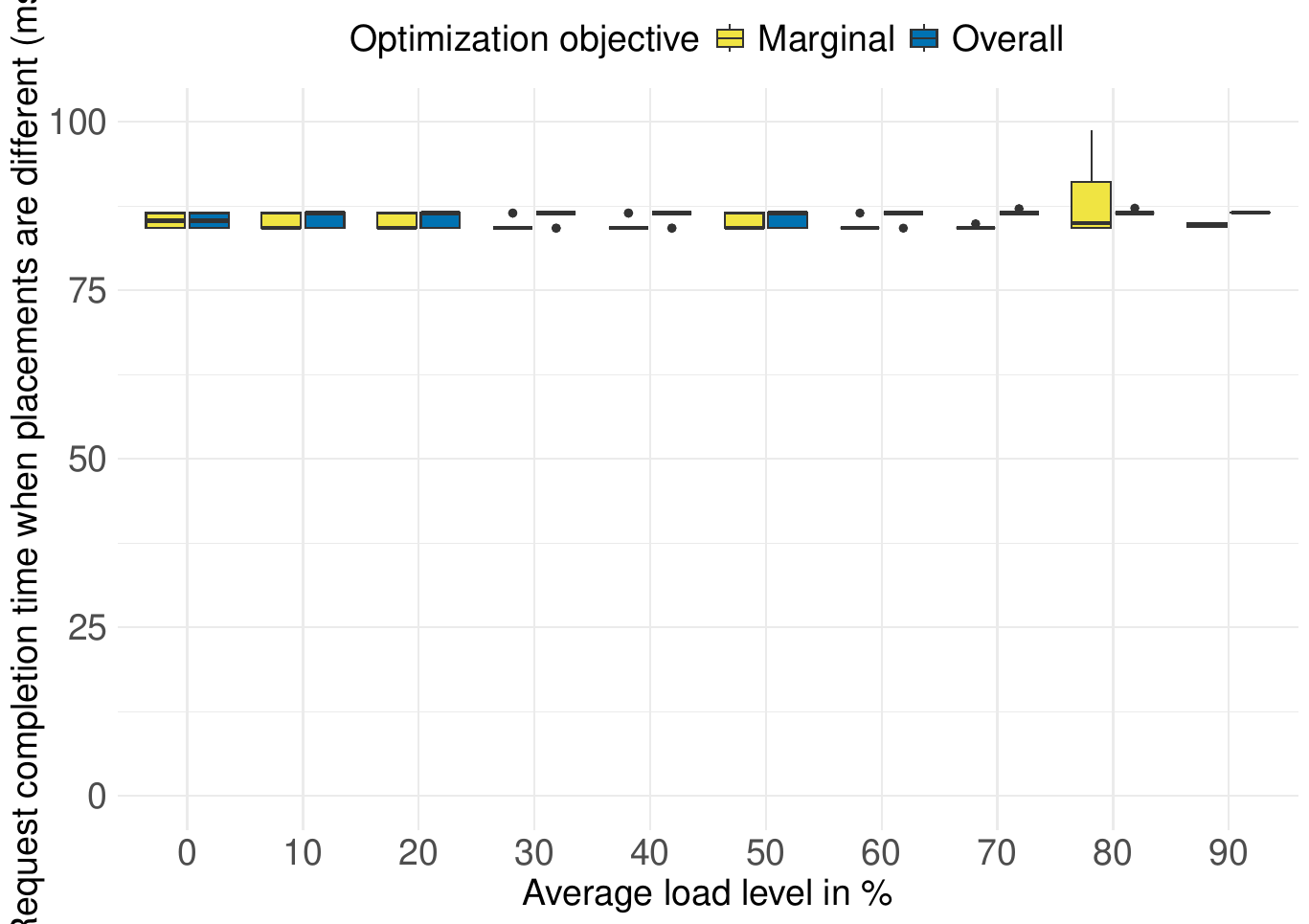}
         \caption{2 avail. function instances per function}
         \label{fig:completionTime2}
     \end{subfigure}
     \hfill
     \begin{subfigure}[b]{0.3\textwidth}
         \centering
         \includegraphics[width=0.9\textwidth]{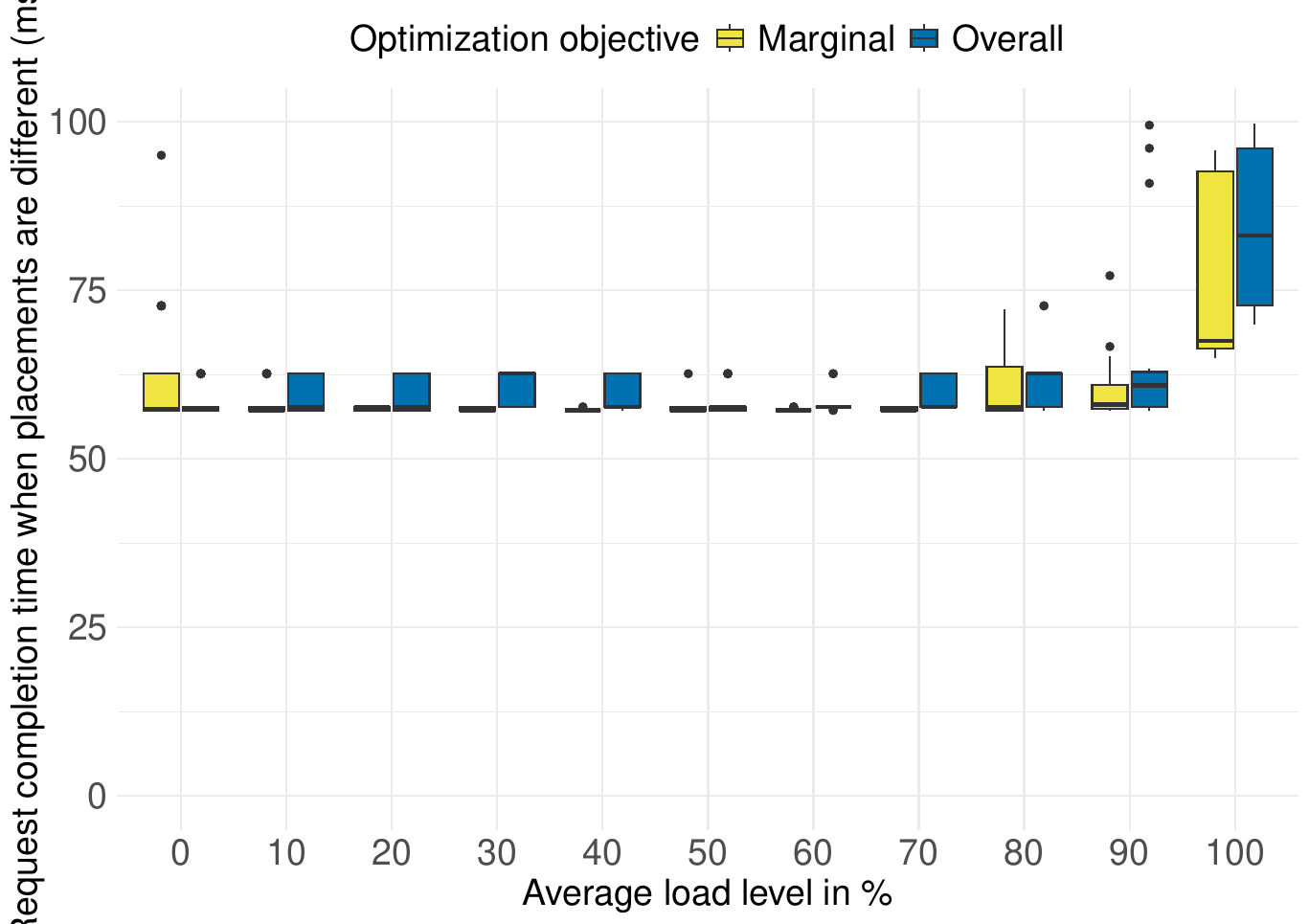}
         \caption{4 avail. function instances per function}
         \label{fig:completionTime4}
     \end{subfigure}
     \hfill
     \begin{subfigure}[b]{0.3\textwidth}
         \centering
         \includegraphics[width=0.9\textwidth]{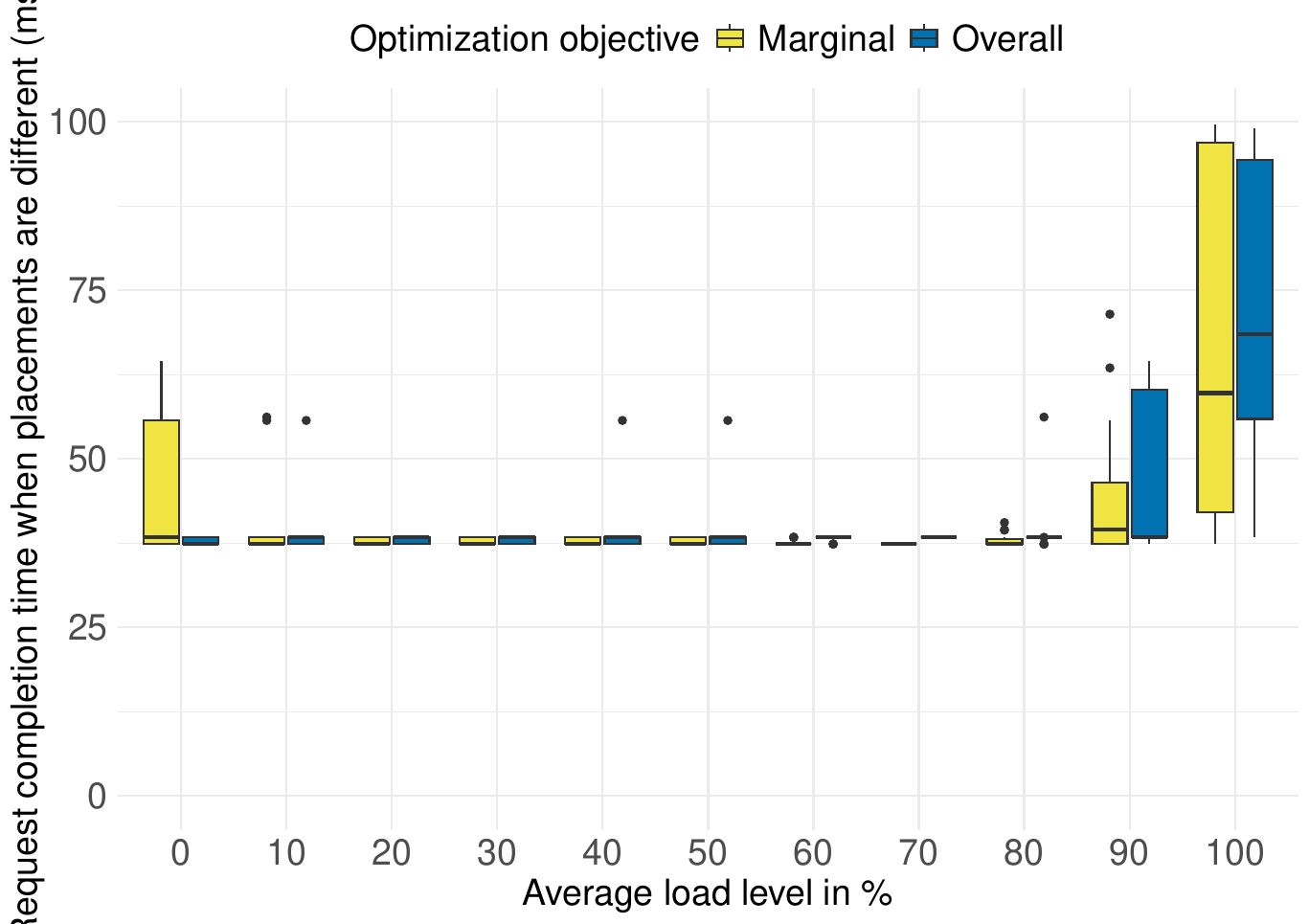}
         \caption{6 avail. function instances per function}
         \label{fig:completionTime6}
     \end{subfigure}
        \caption{Completion time for a given request with varying load and function instance availability.}
        \label{fig:completionTimes}
\end{figure*}

Figure \ref{fig:completionTimes} shows the distribution of the completion time when placing a given request when varying infrastructure load levels and function instance availability. 

\section{Random beginning device study}
\label{ap:RandomRequestingDevice}

\begin{figure*}
     \centering
     \begin{subfigure}[b]{0.3\textwidth}
         \centering
         \includegraphics[width=\textwidth]{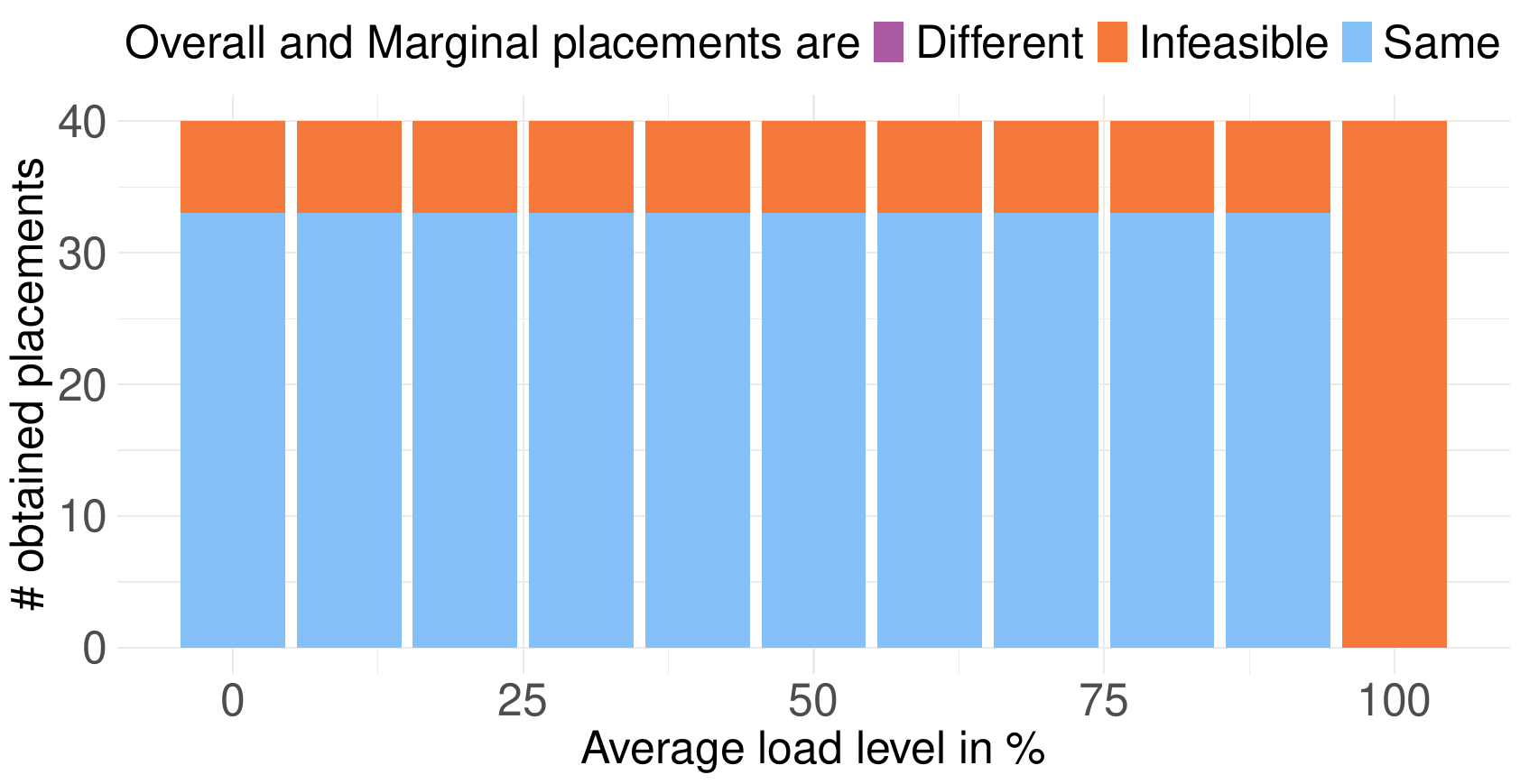}
         \caption{Fixed load at load level}
         \label{fig:loadLevelHomogeneousRandom}
     \end{subfigure}
     \hfill
     \begin{subfigure}[b]{0.3\textwidth}
         \centering
         \includegraphics[width=\textwidth]{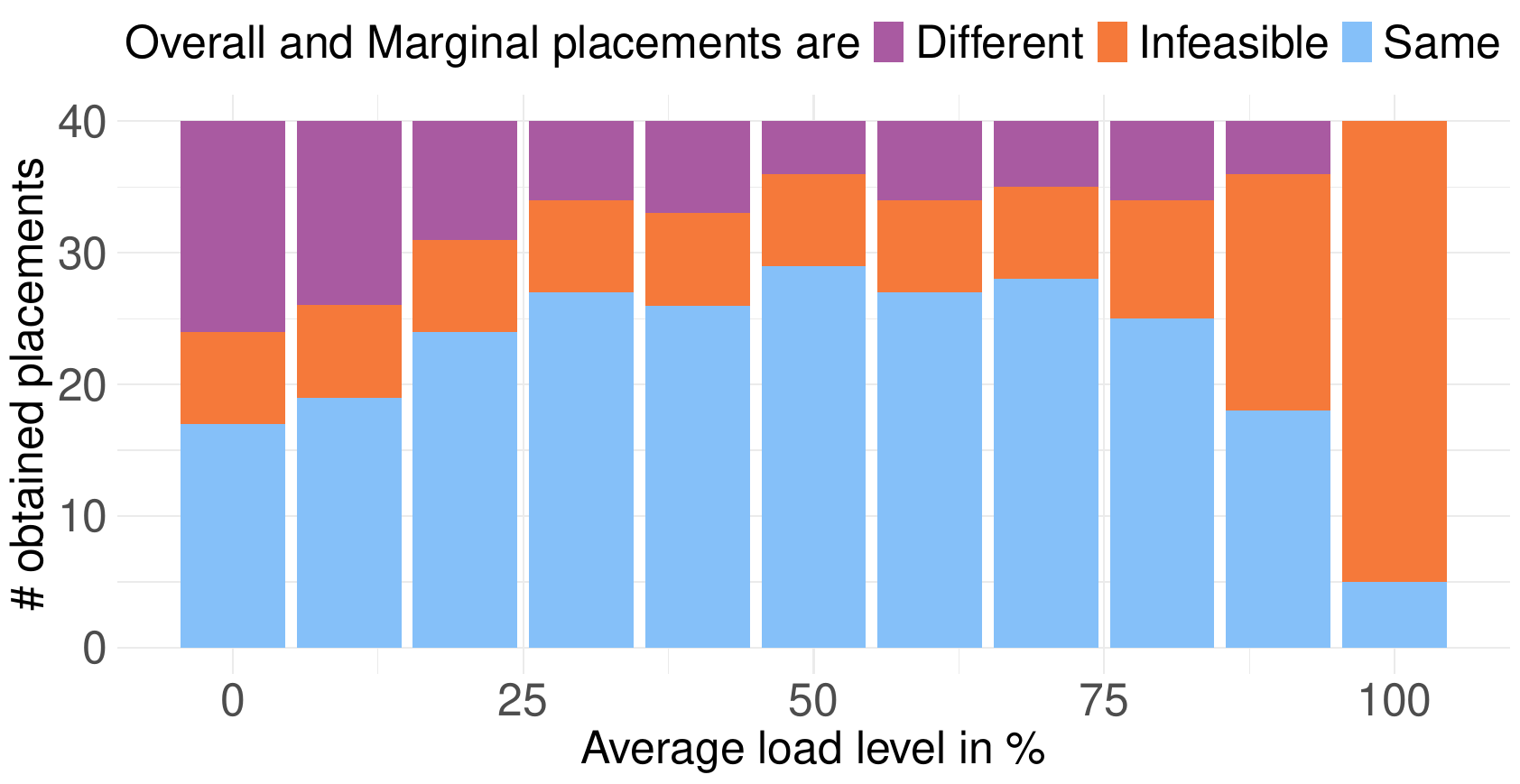}
         \caption{$\mathcal{N}$(load level, 10)}
         \label{fig:loadLevelHeterogeneousRandom}
     \end{subfigure}
     \hfill
     \begin{subfigure}[b]{0.3\textwidth}
         \centering
         \includegraphics[width=\textwidth]{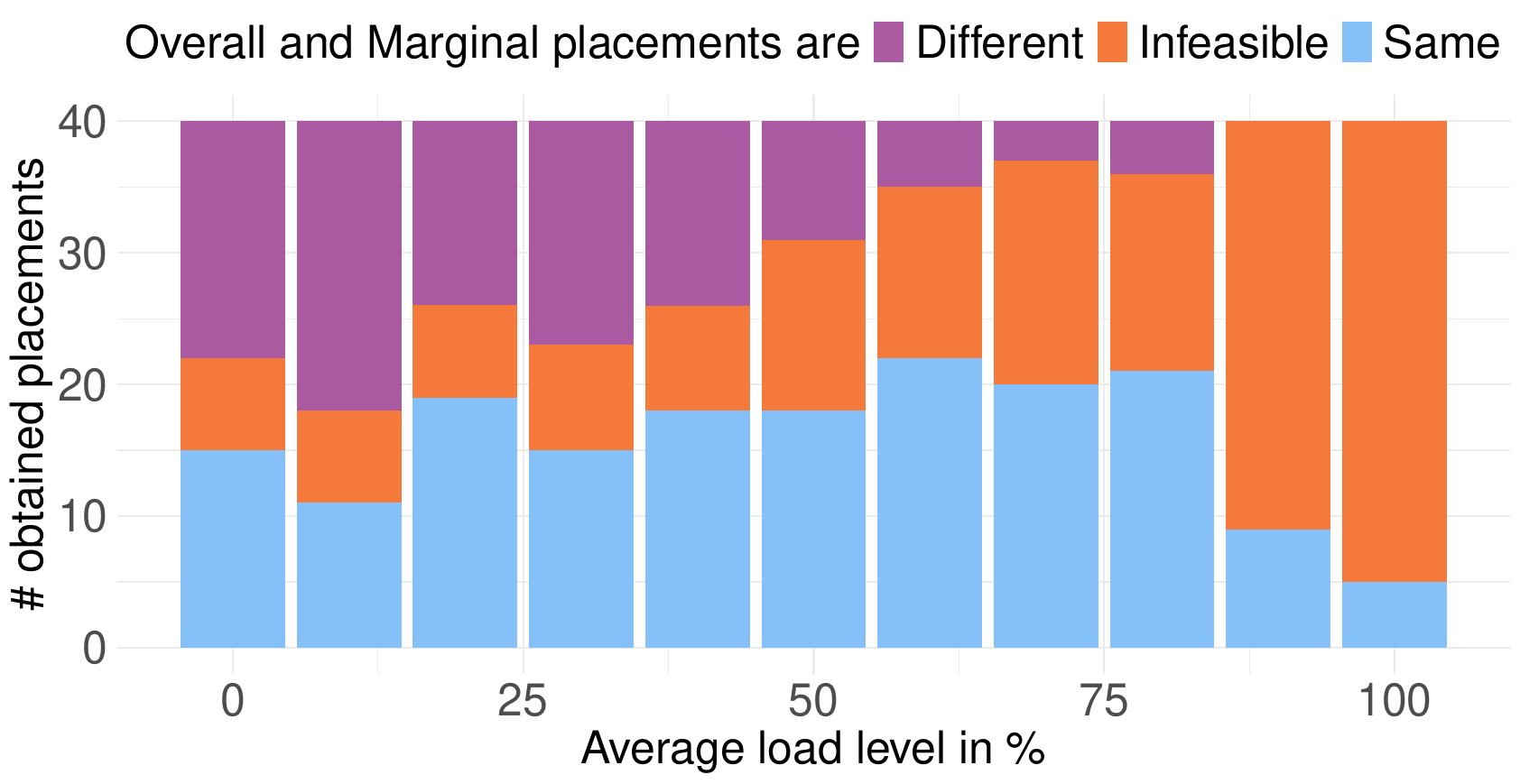}
         \caption{$\mathcal{N}$(load level, 30)}
         \label{fig:loadLevelHeterogeneous30Random}
     \end{subfigure}
        \caption{Categorization of the placements obtained for requests with a random beginning device with different device loads. }
        \label{fig:loadLevelsRandom}
\end{figure*}

\begin{figure*}
     \centering
     \begin{subfigure}[b]{0.3\textwidth}
         \centering
         \includegraphics[width=\textwidth]{images/40rep_DeviceOnlyVar_2replicas_Std10_RequestDeviceRandom_v2.pdf}
         \caption{2 avail. function instances per function}
         \label{fig:functionInstance2Random}
     \end{subfigure}
     \hfill
     \begin{subfigure}[b]{0.3\textwidth}
         \centering
         \includegraphics[width=\textwidth]{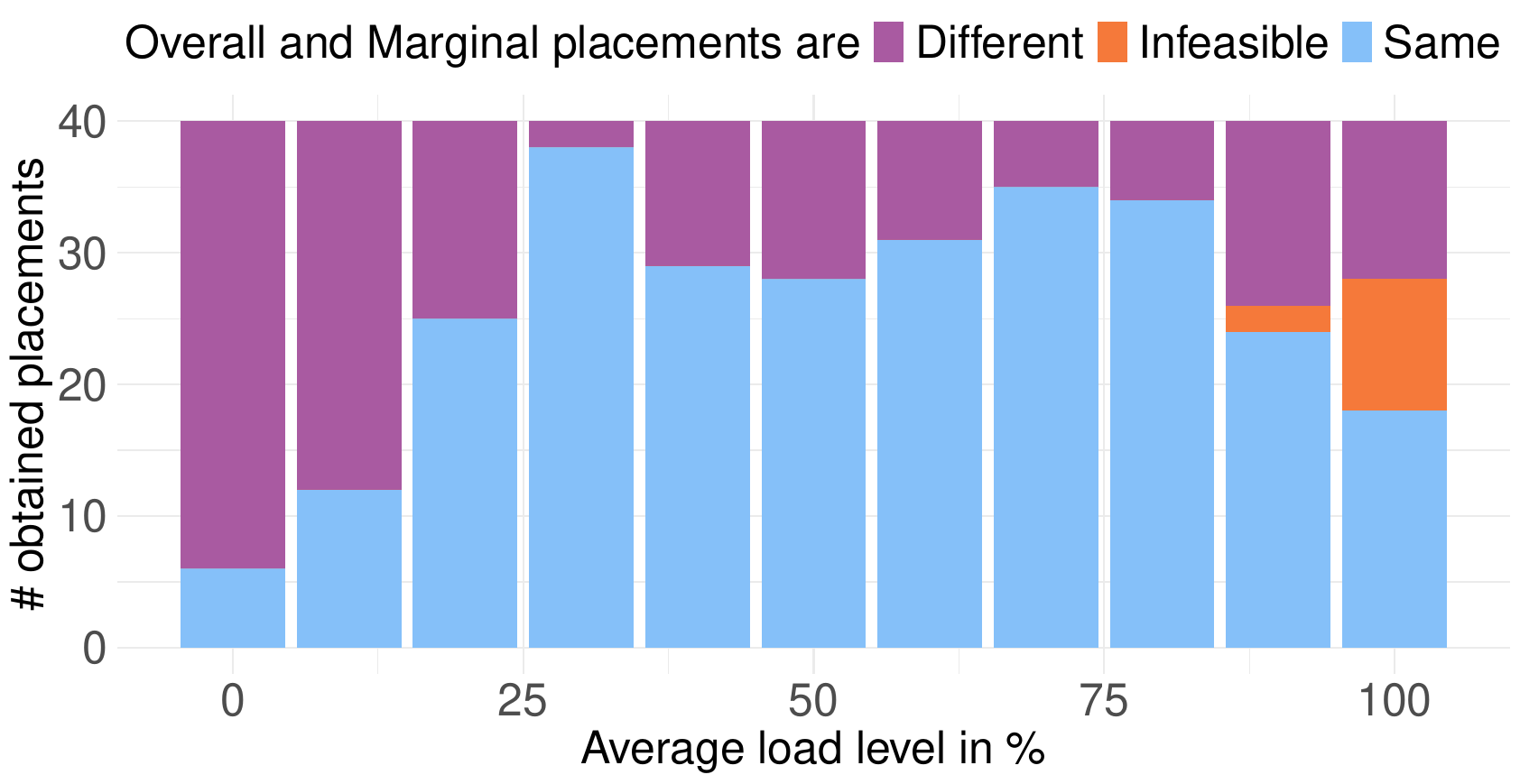}
         \caption{4 avail. function instances per function}
         \label{fig:functionInstance4Random}
     \end{subfigure}
     \hfill
     \begin{subfigure}[b]{0.3\textwidth}
         \centering
         \includegraphics[width=\textwidth]{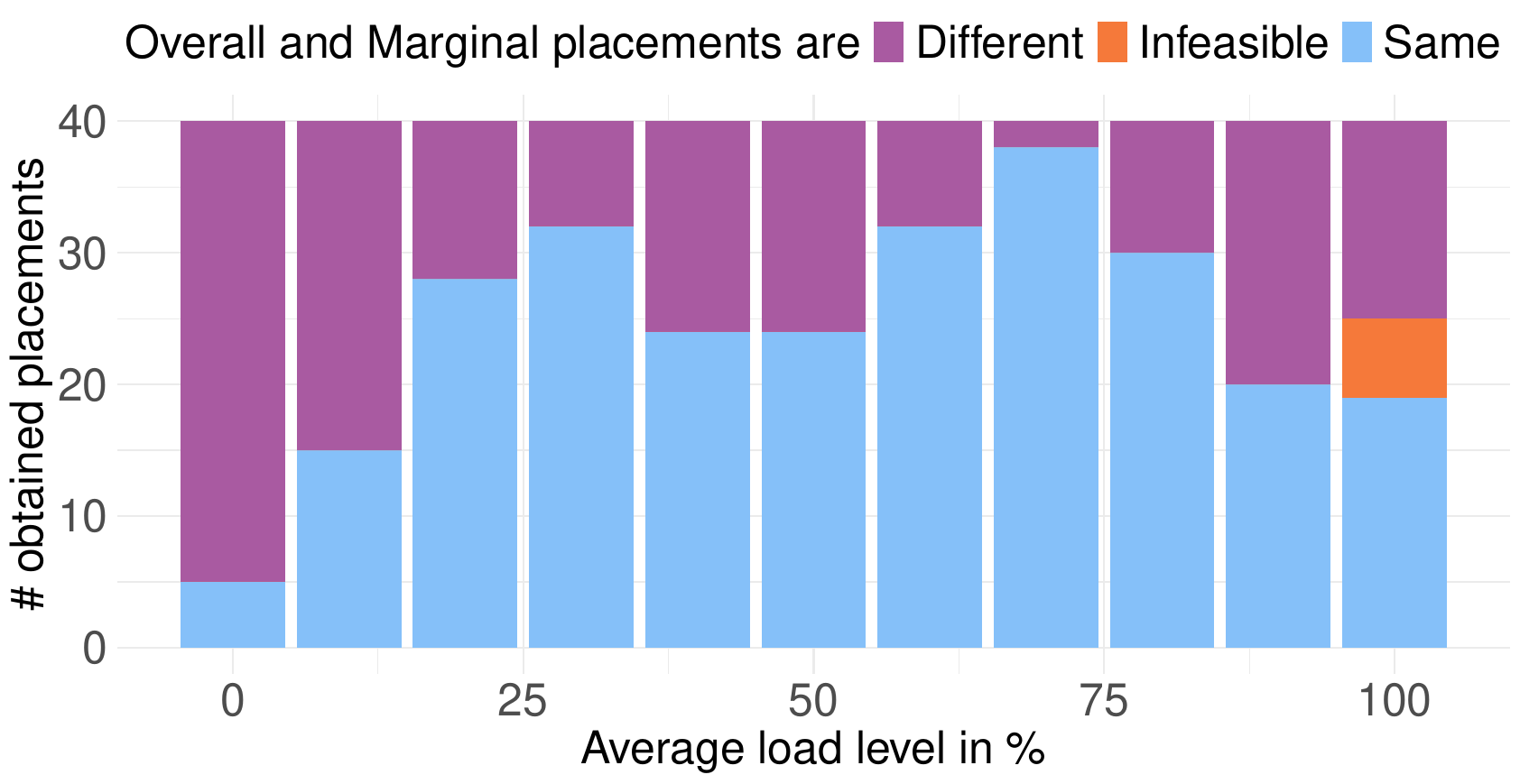}
         \caption{6 avail. function instances per function}
         \label{fig:functionInstance6Random}
     \end{subfigure}
        \caption{Categorization of the placements obtained for requests with a random beginning device with varying load and function instance availability.}
        \label{fig:functionInstancesRandom}
\end{figure*}

\begin{figure*}
     \centering
     \begin{subfigure}[b]{0.3\textwidth}
         \centering
         \includegraphics[width=0.9\textwidth]{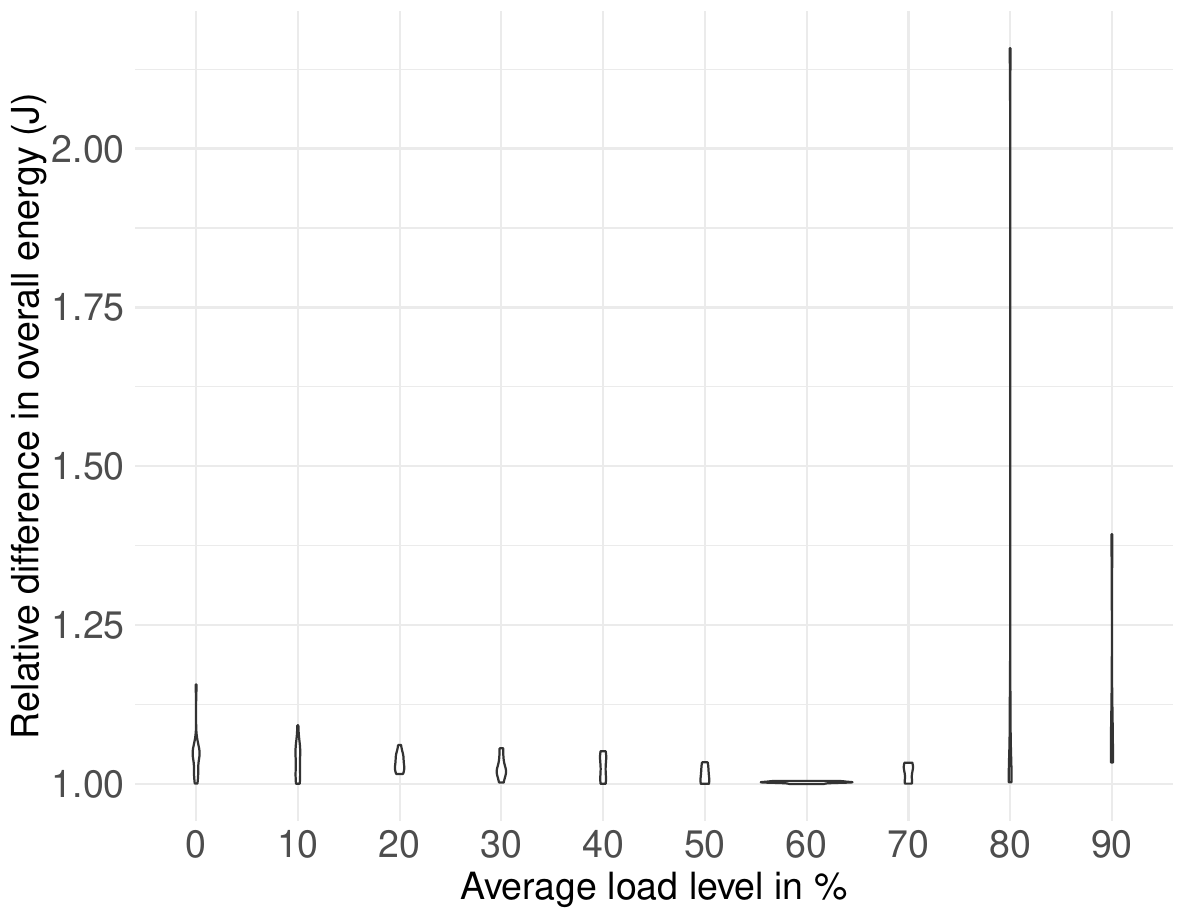}
         \caption{2 avail. function instances per function}
         \label{fig:functionInstance2RelativeDifferenceRandom}
     \end{subfigure}
     \hfill
     \begin{subfigure}[b]{0.3\textwidth}
         \centering
         \includegraphics[width=0.9\textwidth]{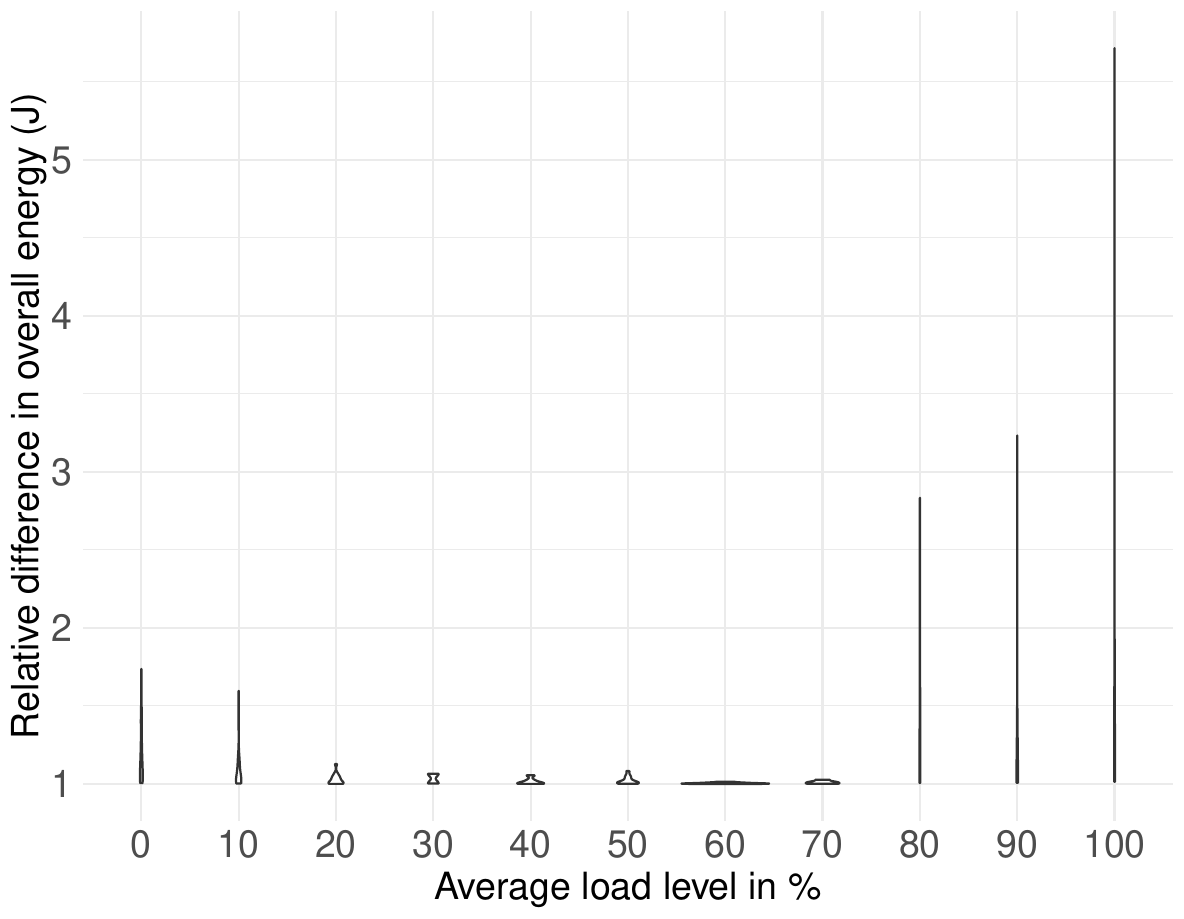}
         \caption{4 avail. function instances per function}
         \label{fig:functionInstance4RelativeDifferenceRandom}
     \end{subfigure}
     \hfill
     \begin{subfigure}[b]{0.3\textwidth}
         \centering
         \includegraphics[width=0.9\textwidth]{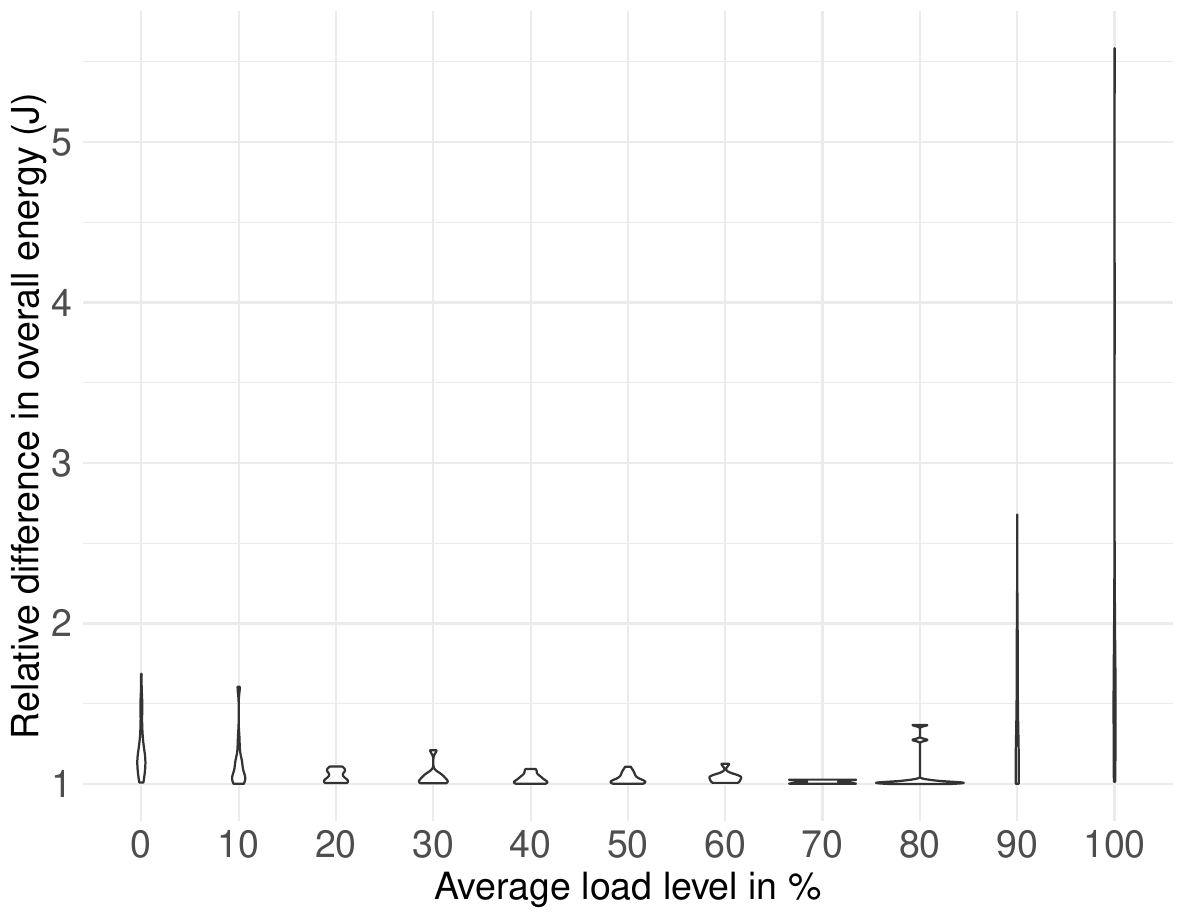}
         \caption{6 avail. function instances per function}
         \label{fig:functionInstance6RelativeDifferenceRandom}
     \end{subfigure}
        \caption{Overall energy relative difference for requests with a random beginning device with varying load and function instance availability (when the resulting placement is different).}
        \label{fig:functionInstancesRelativeDifferenceRandom}
\end{figure*}

\begin{figure*}
     \centering
     \begin{subfigure}[b]{0.3\textwidth}
         \centering
         \includegraphics[width=0.9\textwidth]{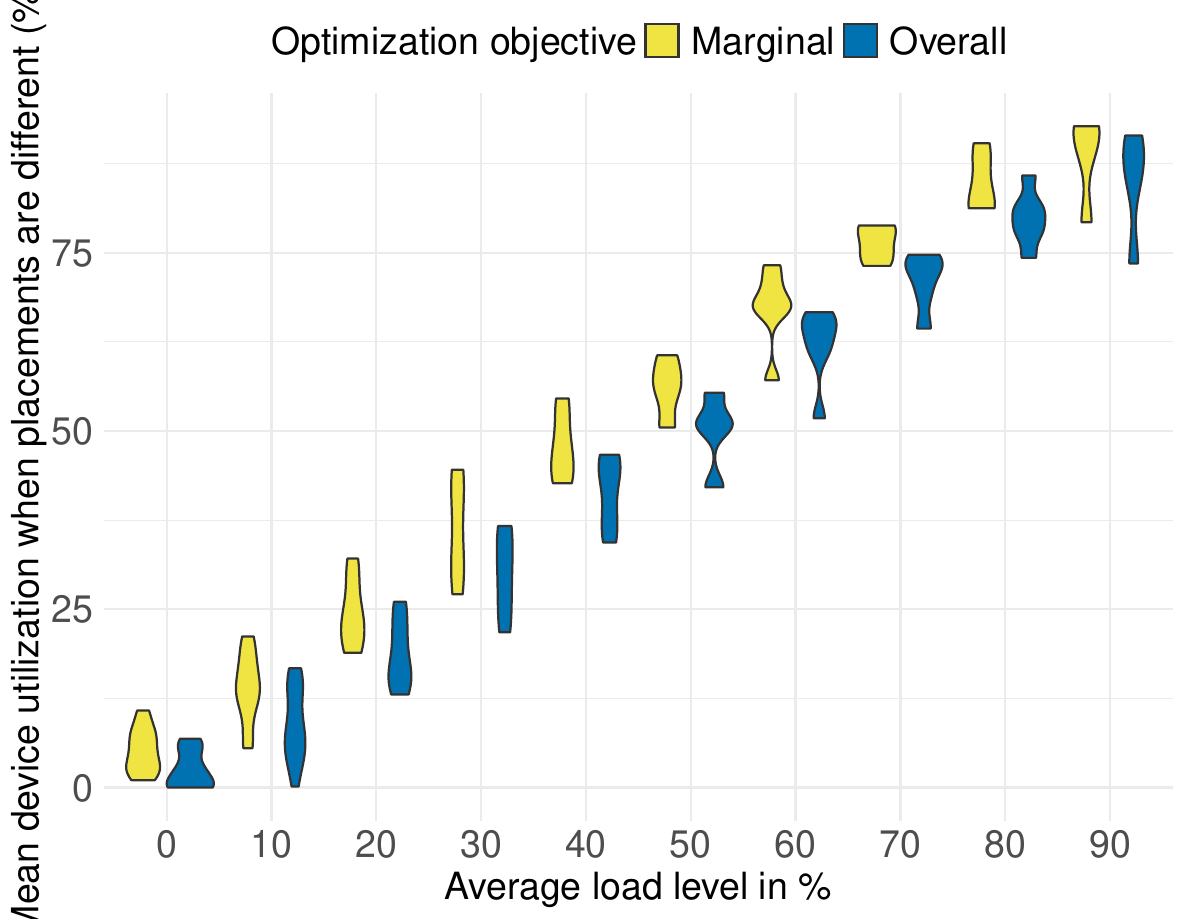}
         \caption{2 avail. function instances per function}
         \label{fig:utilization2Random}
     \end{subfigure}
     \hfill
     \begin{subfigure}[b]{0.3\textwidth}
         \centering
         \includegraphics[width=0.9\textwidth]{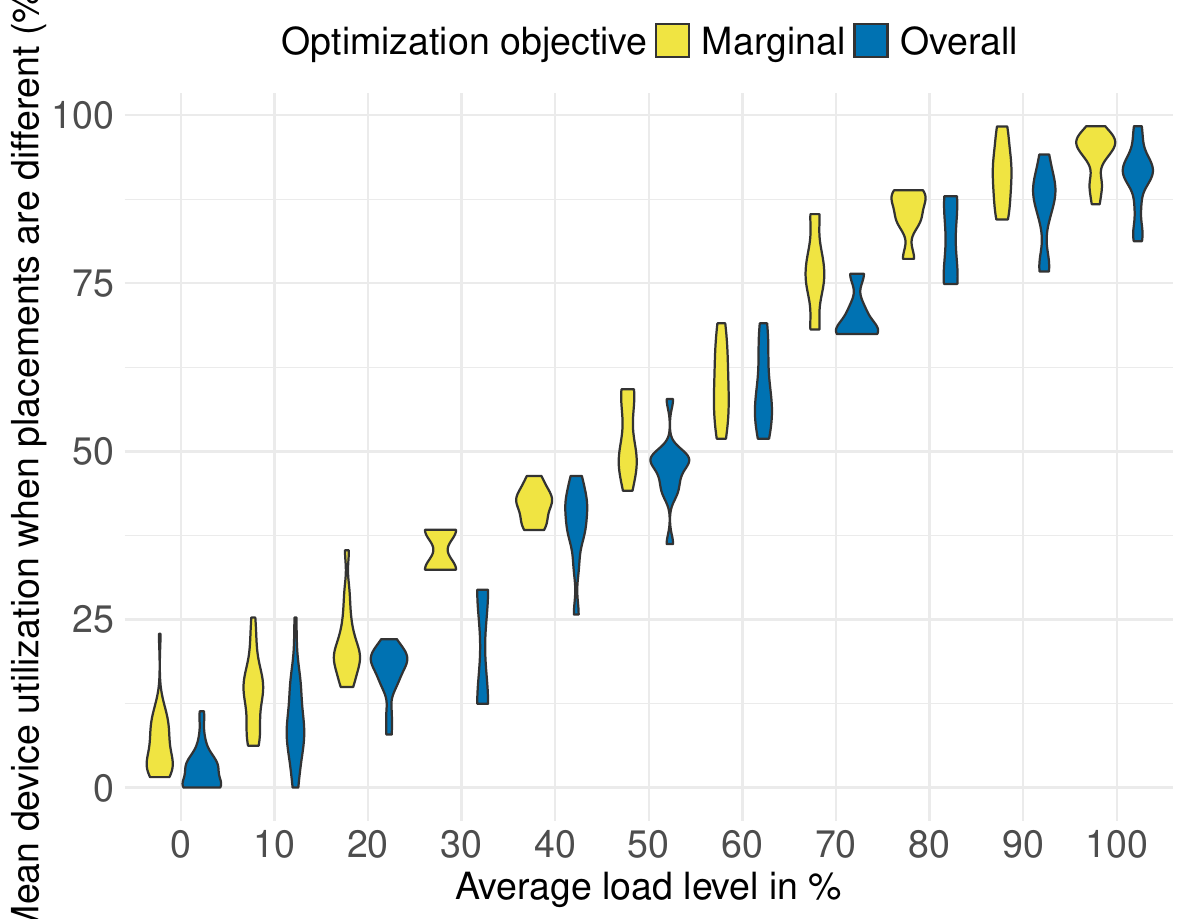}
         \caption{4 avail. function instances per function}
         \label{fig:utilization4Random}
     \end{subfigure}
     \hfill
     \begin{subfigure}[b]{0.3\textwidth}
         \centering
         \includegraphics[width=0.9\textwidth]{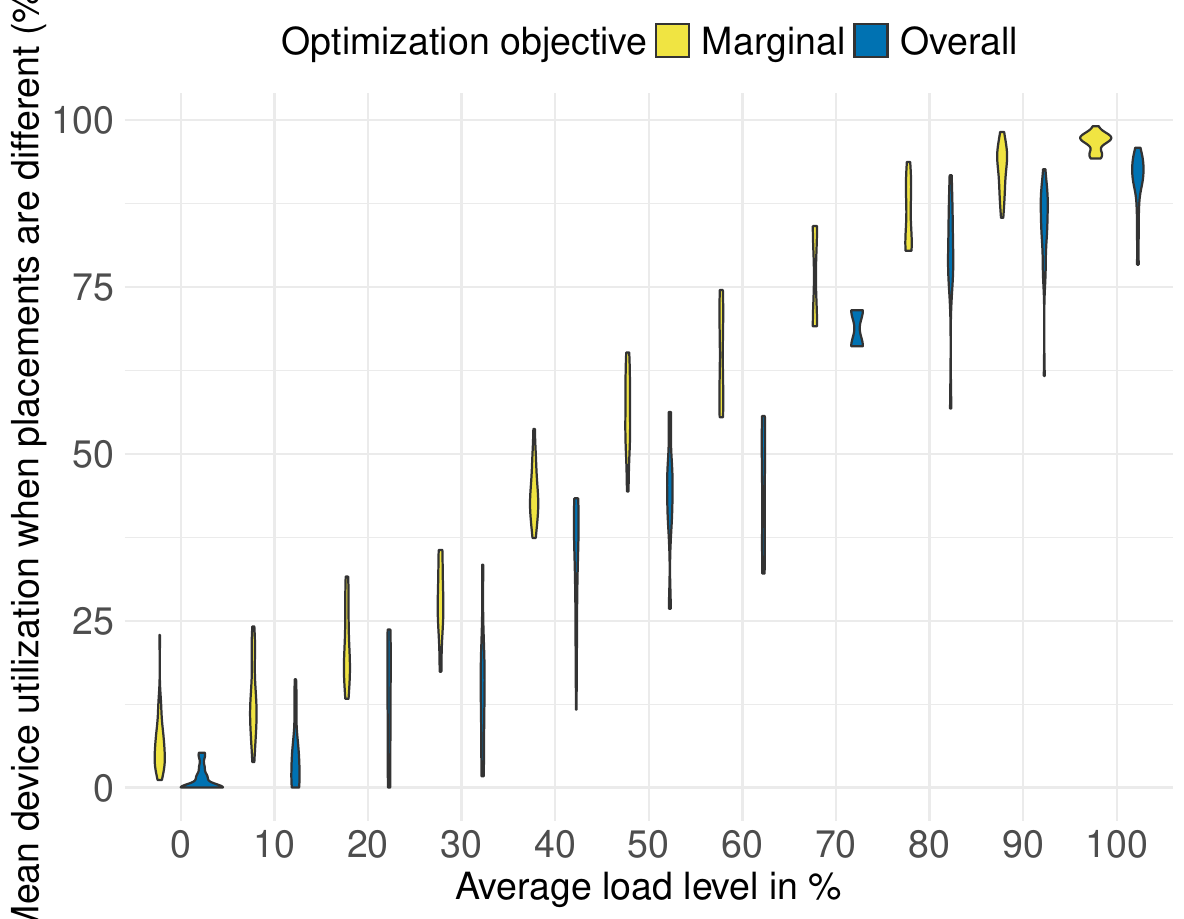}
         \caption{6 avail. function instances per function}
         \label{fig:utilization6Random}
     \end{subfigure}
        \caption{Average device utilization for included devices for requests with random beginning device and with varying load and function instance availability. }
        \label{fig:deviceUtilizationRandom}
\end{figure*}

\begin{figure*}
     \centering
     \begin{subfigure}[b]{0.3\textwidth}
         \centering
         \includegraphics[width=0.9\textwidth]{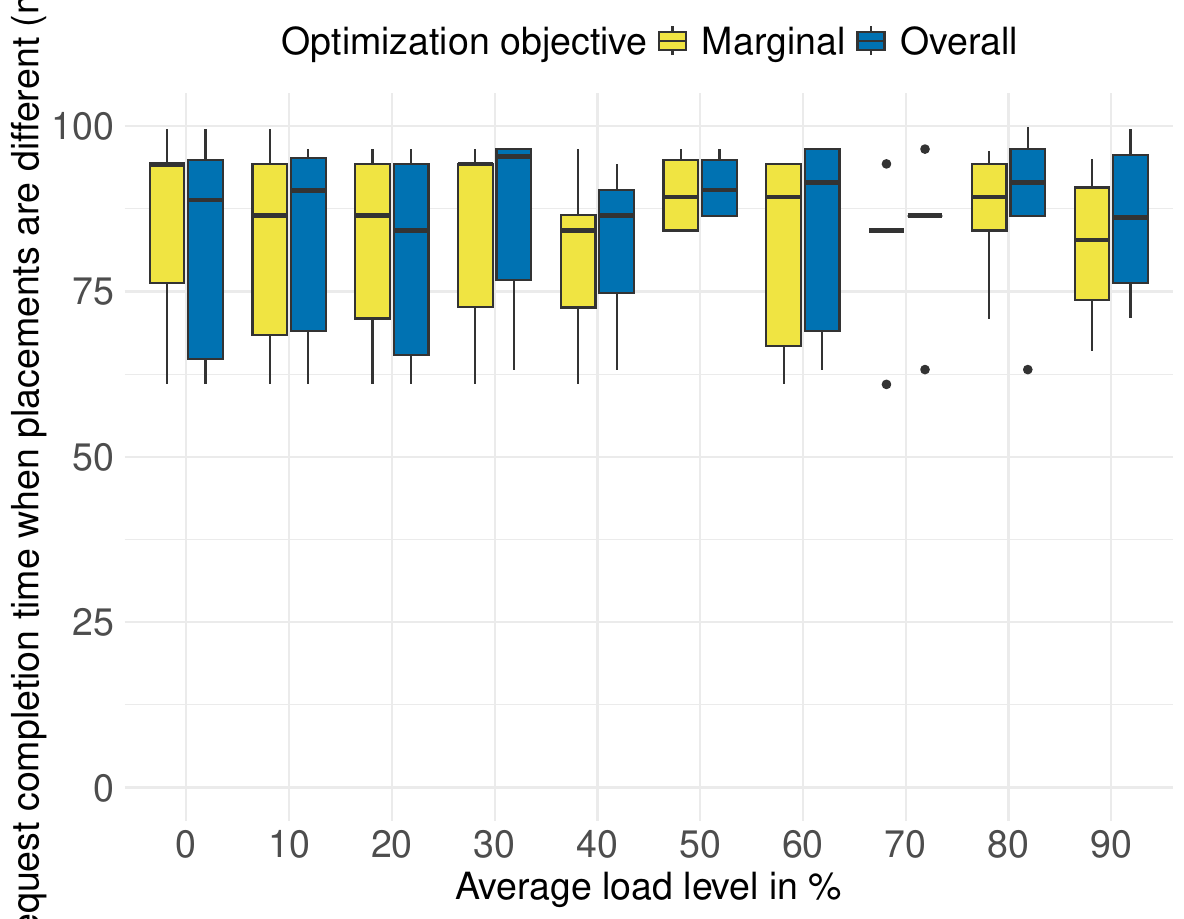}
         \caption{2 avail. function instances per function}
         \label{fig:completionTime2Random}
     \end{subfigure}
     \hfill
     \begin{subfigure}[b]{0.3\textwidth}
         \centering
         \includegraphics[width=0.9\textwidth]{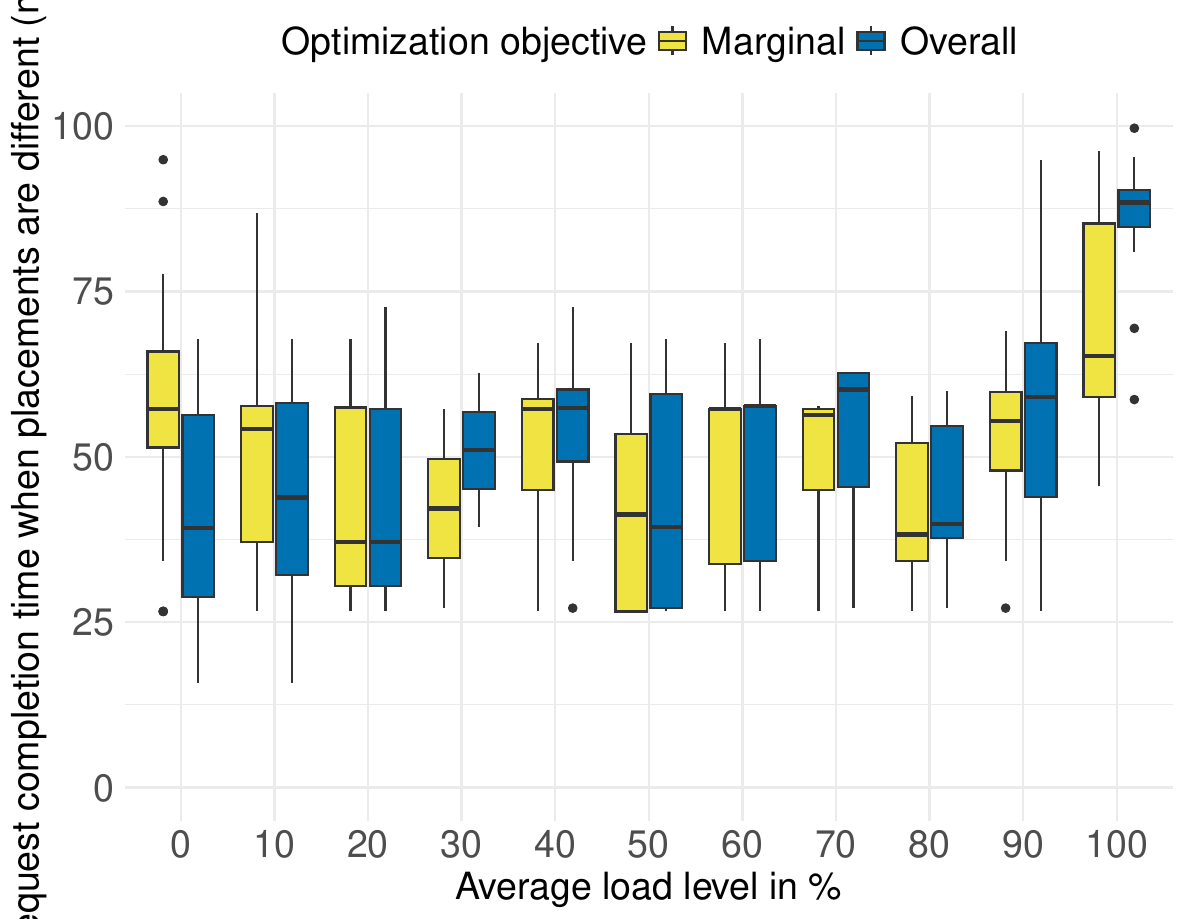}
         \caption{4 avail. function instances per function}
         \label{fig:completionTime4Random}
     \end{subfigure}
     \hfill
     \begin{subfigure}[b]{0.3\textwidth}
         \centering
         \includegraphics[width=0.9\textwidth]{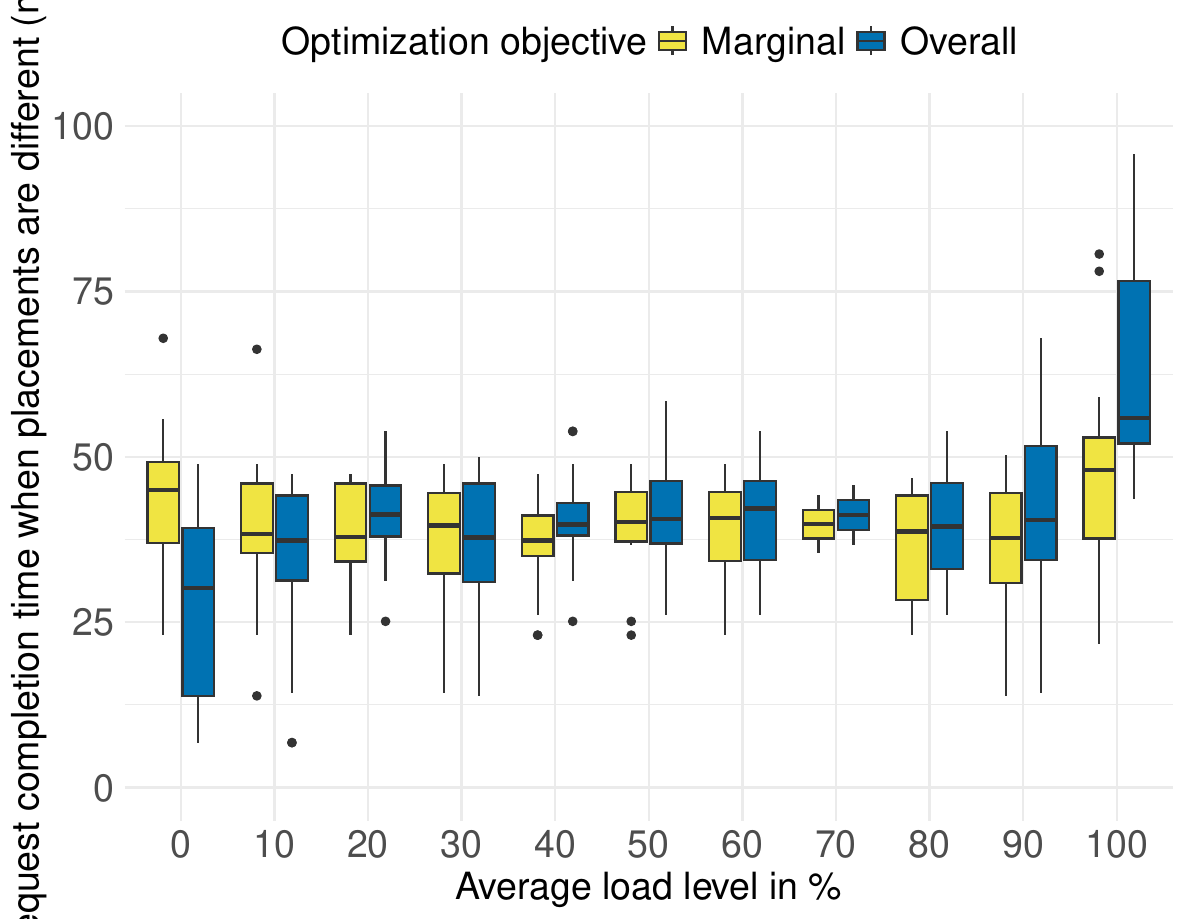}
         \caption{6 avail. function instances per function}
         \label{fig:completionTime6Random}
     \end{subfigure}
        \caption{Completion time for requests with a random beginning device with varying load and function instance availability.}
        \label{fig:completionTimesRandom}
\end{figure*}

Figure \ref{fig:loadLevelsRandom} shows a categorization of the placement obtained when the beginning device is chosen randomly instead of being fixed, for varying levels of load infrastructure heterogeneity.
Figure \ref{fig:functionInstancesRandom} presents a categorization of the placement obtained when the beginning device is chosen randomly instead of being fixed, for varying numbers of available function instances.
Figures \ref{fig:functionInstancesRelativeDifferenceRandom} and \ref{fig:functionInstancesRelativeDifferenceRandom} show the relative difference when placements are different in terms of overall energy, respectively marginal energy. Table \ref{tab:DifferenceExtentRandom} summarizes the corresponding 10th and 90th percentiles.
Figure \ref{fig:deviceUtilizationRandom} shows the distribution of the average device utilization level prior to placement per run. 
Figure \ref{fig:completionTimesRandom} shows the distribution of the request completion times.

\begin{table}[]
    \centering
    \begin{tabular}{|c|c|c|c|c|}
    \hline
        \textbf{Availability} & \multicolumn{2}{c|}{\textbf{Overall energy}}& \multicolumn{2}{c|}{\textbf{Marginal energy}}  \\
        \hline
        \multirow{2}{*}{2 function instances} & 10th & 0.2\% &10th  & 0.8\% \\
        \cline{2-5}
        & 90th  & 6.2\% &90th  & 43.8\% \\
        \hline
\multirow{2}{*}{4 function instances} & 10th  & 0.3\% &10th  & 1.2\% \\
        \cline{2-5}
        & 90th  & 42.7\% &90th  & 108.1\% \\
        \hline
        \multirow{2}{*}{6 function instances} & 10th  & 0.5\% &10th  & 3.0\% \\
        \cline{2-5}
        & 90th  & 53.8\% &90th & 180.9\% \\
        \hline
        
    \end{tabular}
    \caption{Percentiles for the relative difference in energy consumption, for requests with a random beginning device, when placements are different, all load levels.}
    \label{tab:DifferenceExtentRandom}
\end{table}

\begin{figure*}
     \centering
     \begin{subfigure}[b]{0.3\textwidth}
         \centering
         \includegraphics[width=0.9\textwidth]{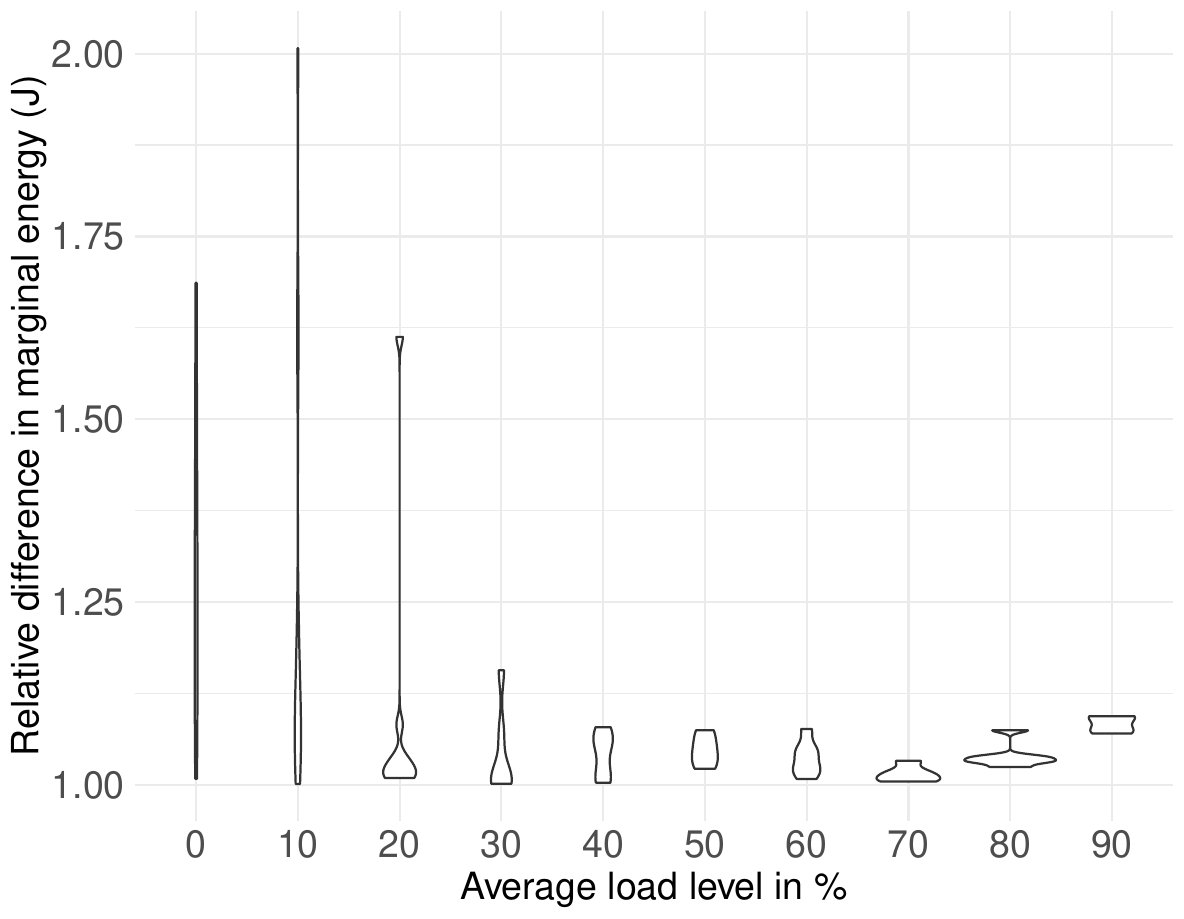}
         \caption{2 avail. function instances per function}
         \label{fig:functionInstance2RelativeDifferenceMargRandom}
     \end{subfigure}
     \hfill
     \begin{subfigure}[b]{0.3\textwidth}
         \centering
         \includegraphics[width=0.9\textwidth]{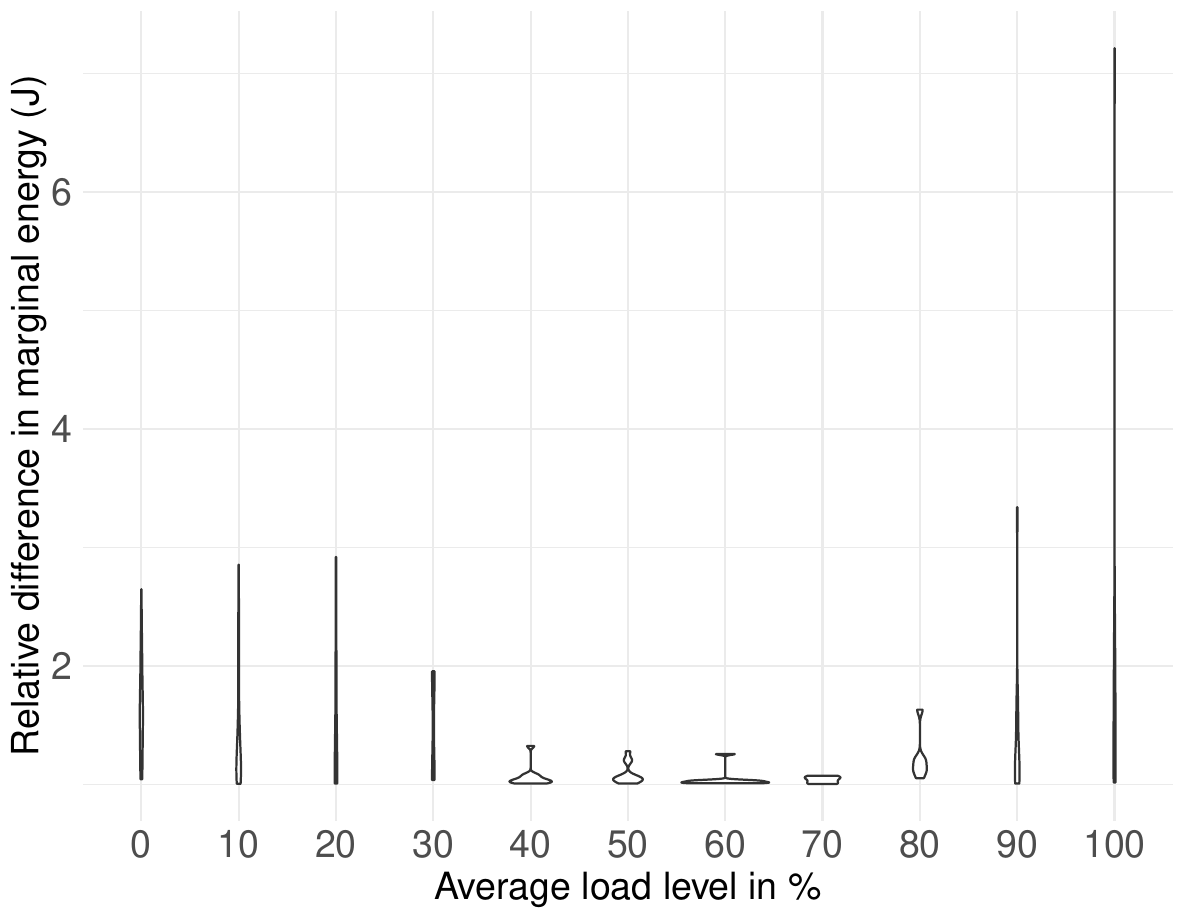}
         \caption{4 avail. function instances per function}
         \label{fig:functionInstance4RelativeDifferenceMargRandom}
     \end{subfigure}
     \hfill
     \begin{subfigure}[b]{0.3\textwidth}
         \centering
         \includegraphics[width=0.9\textwidth]{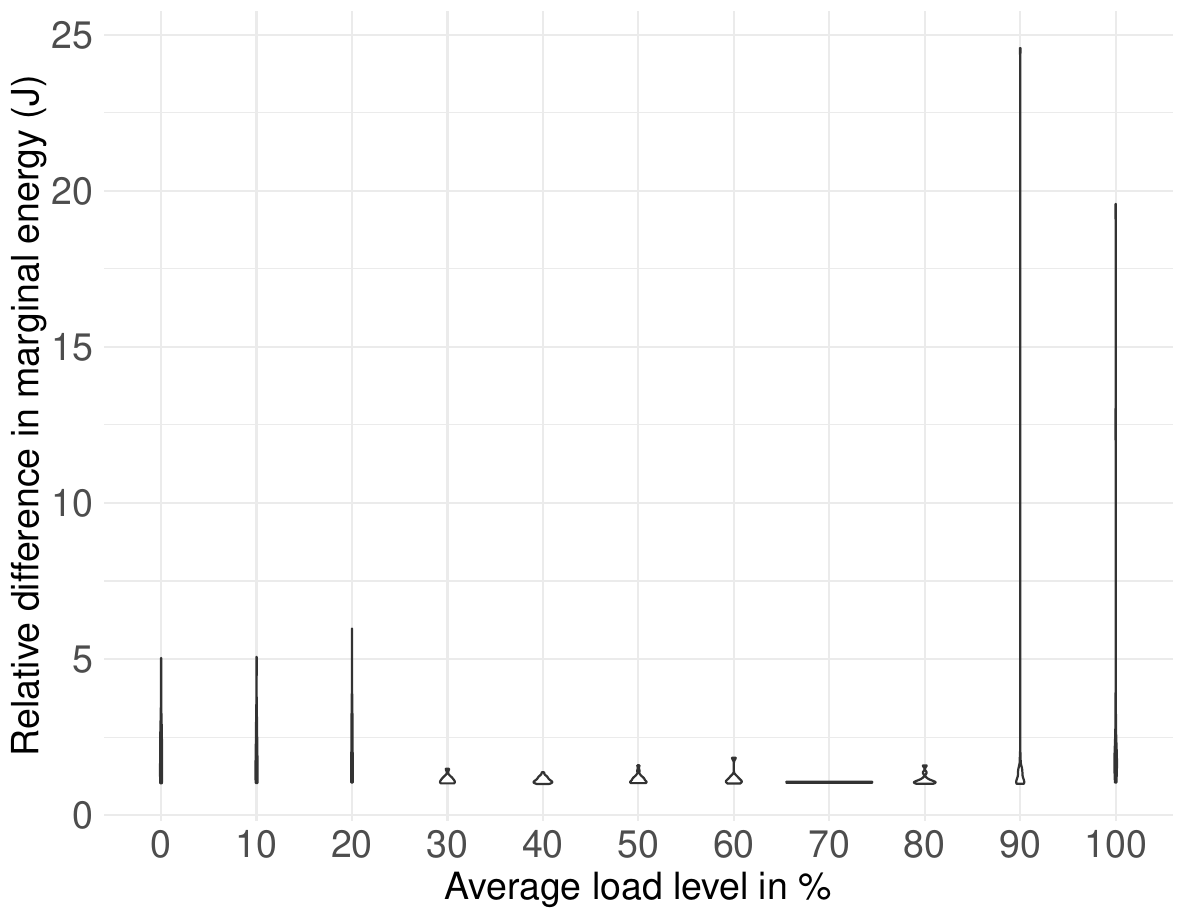}
         \caption{6 avail. function instances per function}
         \label{fig:functionInstance6RelativeDifferenceMargRandom}
     \end{subfigure}
        \caption{Marginal energy relative difference for requests with a random beginning device with varying load and function instance availability (when the resulting placement is different).}
        \label{fig:functionInstancesRelativeDifferenceMargRandom}
\end{figure*}

\end{document}